\documentclass{report}
\pdfoutput=1
\usepackage[a4paper, margin=2.5cm]{geometry}

\author{Luke Storry}
\title{augKlimb:\\ Interactive Data-Led Augmentation of Bouldering Training}
\date{June 2019}
                
\usepackage{pdfpages}
\usepackage{hyperref}
\usepackage{graphicx}

\begin{document}

\maketitle

\chapter*{Abstract}

\noindent
Climbing is a popular and growing sport, especially indoors, where climbers can train on man-made routes using artificial holds.
Both strength and good technique is required to successfully reach the top of a climb, and often coaches work to improve technique so less strength is required, enabling a climber to ascent more difficult climbs.
Various aspects of adding computer-interaction to climbing have been studied in recent years, but there is a large space for research into lightweight tools to aid recreational intermediate climbers, both with trickier climbs and to improve their own technique.

In this project, I explored which form of data-capture and output-features could improve a climber's training, and analysed how climbers responded to viewing their data throughout a climbing session, then conducted a user-centred design to build a lightweight mobile application for intermediate climbers.

A variety of hardware and software solutions were explored, tested and developed through  series of surveys, discussions, wizard-of-oz studies and prototyping, resulting in a system that most closely meets the needs of local indoor boulderers given the project's time scope.

\noindent
\begin{itemize}
\item I spent over $60$ hours conducting in-field observations of climbers interacting with various prototypes.
\item I iteratively developed an interactive mobile app that: \begin{itemize}
      \item can record, graph, and score the acceleration of a climber, as both a training tool and gamification incentive for good technique
      \item can link a video recording to the acceleration graph, to enable frame-by-frame inspection of weaknesses
      \item is fully approved and distributed on the Google play Store and currently being regularly used by 15 local climbers.
\end{itemize}
\item I wrote over $1000$ lines of \verb|C#| source code, with a further $20,000$ lines of Unity code-files defining the graphical interface.
\item I conducted a final usability study, comprising a thematic analysis of forty minutes's worth of interview transcripts, to gain a deep understanding of the app's impact on the climbers using it, along with its benefits and limitations. 
\end{itemize}
\chapter*{Acknowledgements}
Firstly, a massive thanks must go to my dissertation supervisor Peter Bennett, for both aiding me greatly in the formation of the initial idea, and providing invaluable support throughout the execution of the project.
I want to thank Beth, the department's Teaching Technologist, for lending me a second android phone, which greatly sped up development of the app.
To the three climbers who lent me hours of their time to test various iterations of the app, and the six climbers who enabled me to perform my final in-depth analysis: thank you!
And of course my unending gratitude to the Bristol climbing community, for providing responses to my surveys, performing countless beta-tests, giving me the much-needed feedback I needed to develop the app.

\chapter*{Supporting Technologies}

\begin{itemize}
\item I used Unity to develop my app, as its simple UI builder, support for multiple mobile platforms and rich library of Asset addons made developing and iterating my designs much quicker and easier. \url{https://unity.com/}
\item I used the free NativeCamera (\url{https://assetstore.unity.com/packages/tools/integration/native-camera-for-android-ios-117802}), NativeShare (\url{https://assetstore.unity.com/packages/tools/integration/native-share-for-android-ios-112731}), NativeGallery (\url{https://assetstore.unity.com/packages/tools/integration/native-gallery-for-android-ios-112630}), and SimpleFileBrowser (\url{https://assetstore.unity.com/packages/tools/gui/runtime-file-browser-113006}) Unity Assets to add those features to my app without wasting time developing the code myself.
\item I used parts of the OpenCV Computer Vision Library to explore frame-by-frame video-analysis, mainly edge-detection and blob-detection.
\\\url{https://opencv.org/}
\\\url{https://assetstore.unity.com/packages/tools/integration/opencv-plus-unity-85928}
\item I used an Android Mobile device supplied by the Department, as well as my own Android device, to help build and debug the app during development.

\end{itemize}

\tableofcontents

\chapter{Contextual Background}
\label{chap:context}

%

In this chapter, I aim to outline what bouldering is, why recreational intermediate boulderers were my target audience and test-population, the gaps in literature that this project aims to explore.
I also detail the four hypotheses that guided the development of both the app and the study, and outline some of the key challenges this project faced.

\section{The Problem Being Investigated}
\subsection{Indoor Bouldering}
\subsubsection{What}

For the purpose of this project, I will be studying how interacting with data affects recreational indoor boulderers.
Bouldering is a sub-genre of climbing where the routes are generally quite short (under 5m), so no ropes or gear are required, only a soft crash-mat for safety.
\begin{figure}[h]
\centering
\includegraphics[width=4cm]{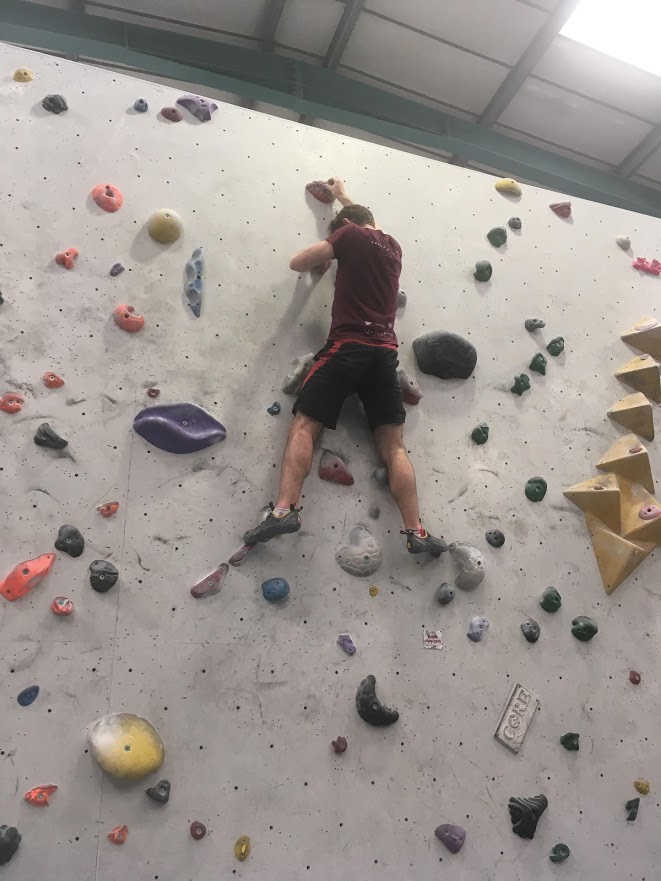}
\includegraphics[width=4cm]{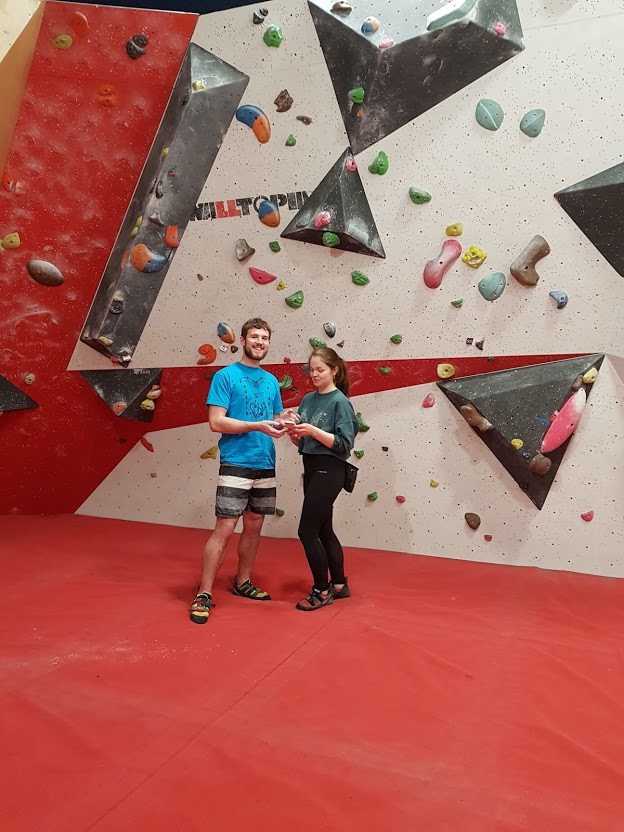}
\caption{Example images of a bouldering wall}
\label{fig:playstore}
\end{figure}

Many indoor bouldering ``gyms" have opened in the past decade, with the low requirements for kit and more social environment being mentioned as some reasons for this growth~\cite{socialclimb}.

\subsubsection{Why}
Studying an indoor variant of climbing was chosen as it is weather in-dependant (a useful factor for a studying being conducted in the winter-spring seasons).
Also, it was a lot more accessible for both me and my potential testers, with three locations across Bristol, all within easy cycling distance of the university.
This meant that many more test sessions could take place, between two and four three-hour sessions per week, which is much more than would have been possible with a drive out to a nearby rock-face.

Another reason indoor climbing (and the bouldering sub-type in particular) was chosen is because it usually involves shorter, slightly harder routes, with participants often discussing the climbs between attempts.
In addition, the shorter routes are easily visible from the ground, with fairly consistent lighting, potentially allowing a stationary camera-equipped device to observe the entire climb in-frame.

\subsubsection{Grades}
As a sport, the goal of climbing is to ascend (go up) or traverse (go across) a wall, from the ground to an end-point.

Some of these routes are more difficult than others, and so they can be ``graded" in various ways to indicate to other climbers the relative difficulty.
This difficulty is generally subjective, and is usually a combination of: the type and convenience of the holds, the distance between holds, and the incline of the wall.
Some people use this as just an indicator for which climbs they are likely to have the most fun on, and other use this as a tool for measuring their progression, trying to climb the highest grade they can.
Climbing routes that are easier than a persons maximum ability offers an opportunity to refine and hone strength and technique before attempting harder-graded climbs.

There are a variety of ways of grading routes.
Often bouldering uses a ``$V$" system, with $V0$ being easy beginner-orientated climbs, and $V7$ being an expert-level route.
In the old outdoor bouldering, this was determined by discussion and opinions on various natural holes in rocks, but in modern indoor bouldering gyms, climbing routes are designed, or ``set" with all the hand- and foot-holds being made of a certain coloured plastic, and that colour will signify it's difficulty.

\subsection{Intermediate-level Climbers}
To further help narrow the scope of the project, I decided to aim a potential product or interactive device on a typical intermediate climber.
This is because beginners often progress very quickly with just \textit{more time} spent climbing, and expert-level climbers often have their own coaches to aid in their growth, whilst a large number of intermediate climbers, who want to improve, but may not know what they need to work on to improve, and get no feedback from their current climbing.

For many intermediate boulderers who are facing a plateau in their progression, they currently face two options: either to continue climbing a lot at a low grade, hoping they gradually build up the strength and technique required to break through and climb harder; or to pay (quite a lot of) money for a private coach who can analyse their technique and give them precise feedback on how to improve.

Therefore from a climbing-improvement perspective, the product I was aiming to build could either incentivise climbing lower-grade with a high-volume, or give feedback on technique, or both.

\section{Importance of Topic}


There is currently no lightweight device, product or app in the market that is aimed at helping intermediate climbers see data and progress in their climbing.
As will be further discussed in Chapter~\ref{chap:technical}, academic research has thus-far focused on very specialised hardware or custom-built climbing-walls, with little attention on something smaller or lightweight that is highly-accessible for every climber to use.

In conducting a user-centred iterative design to discover what kind of data interaction is most useful for intermediate climbers, I have both built a product that is fit-for-purpose, and have begun exploring what is an as-of-yet under-studied area in Human-Computer Interaction:
In-field user studies that aim to capture the view and use-patterns of the majority of boulderers.

\section{Hypotheses}
Along with developing a product, from a HCI perspective I also wanted to see how climbers interact with a more modern product or app.
Many climbers either ``just climb", leaving their phone in the lockers, or if they do track it they simply list which climbs they have completed in a logbook or logging-app, and there has been little research into the computerisation of mainstream climbing aids thus far.
Does the adding of some measurements and metrics aid a climber in the long run,
either by making climbing more fun and therefore causing them to go train more often, or does the frequency not change as much as the effectiveness of the training itself - instead of aimlessly climbing with a vague goal of ``getting better", does providing quantitative analysis of the climbs completed give more of a focus to the sessions that a climber does?

With these thoughts in mind, I formulated four key hypotheses to explore:
\begin{enumerate}
    \item Augmenting a climbing session with a live-feed of data analytics will positively impact climbing technique.
    \item A lightweight, low-cost and simple-to-use product will be popular among intermediate climbers who are serious enough to want to improve, but not so serious they want to pay for coaching.
    \item Seeing a ``score" that rates climbing technique will enable gamification and fun, both for individuals and within groups.
    \item Providing more data to climbers will enable more focused progression tracking and goal-oriented training.
\end{enumerate}

\section{Central Challenges}
One central challenge of the project was to iteratively develop a useful product that fits the needs of local indoor intermediate boulderers.
The associated second central challenge was the required testing, analysis and understanding of how the augmentation of live data analytics can aid a climber's progress or enjoyment.

With these initial goals, after months of study and through four major iteration-cycles, I developed an app that can record, share, and match accelerometer data and video recordings taken from a variety of devices at different times.
This was the third major challenge - alongside designing a product and investigating its impact on users, actually coding and publishing a feature-rich app that is reliable and fast on a range of devices was a big challenge.

Sub-challenges included:
\begin{enumerate}
\item designing a series of metrics that are useful and consistently comparable between climbs, \item exploring the video-analysis of climbing technique,
\item constant user testing and re-evaluation of goals, quickly developing features in time for a thrice-weekly in-field testing session at the wall,
\item ethical considerations of developing a product that advises people whilst they perform an inherently risky sport - both during the development of ideas, and the two full academic-ethics applications
\item running a series of longer-term user-analyses and interviews at the end of the project, to further explore the impact of using the completed app, then conducting a thematic analysis over the transcripts of the interviews.
\end{enumerate}

\chapter{Technical Background}
\label{chap:technical}

\section{Previous Sports-Data-Augmentation Studies}
\subsection{Impact of Feedback on Sport Progression}
The impact of appropriate feedback on motor skill acquisition has been widely studied in the field of psycho-kinesiology~\cite{schmidt75aschema, schmidt2005motor}.
One of the earliest reviews of the impact that more modern computer-based feedback could have on athlete's performance is Liebermann et al's \textit{``Advances in the application of information technology
to sport performance"}~\cite{lieberreview}, which discusses how video-based review systems, eye-tracking technology, and force-plates could all be used by coaches to provide an athlete with ``sophisticated and objective" feedback during a training session, leading to ``effective and efficient learning" across a wide variety of sports.

Although one of the aims of the project was to not require a coach, the clearly-defined link between high-quality feedback and learning speed that was presented by those two papers was promising, especially paired with the many studies contained within that review that showed the efficacy of computer-aided feedback systems.

In a their study~\cite{bacafeedback}, Baca \& Kornfeind examined the use of advanced computerised ``Rapid Feedback Systems" in elite training for rowing, table-tennis and biathalon, resulting in a set of considerations and guidelines for when designing such a system.
Among others, these included:
\begin{itemize}
    \item minimising the impact that the measurement system has on the athlete
    \item ensuring the system is mobile, so usable at the training location, to prevent restriction to laboratory environments
    \item providing a fast, comprehensible and easily-decipherable GUI
    \item reducing the setup and training time required for proficient usage of the system.
\end{itemize}

Again, this study included the interaction of coaches, but it provided a useful discussion on the efficacy of instant easy-to-use feedback systems, along with these guidelines for developing my product going forwards.

\subsection{Video}
Using video-playback to aid in the analysis of sports performance is a common traditional coaching technique, with studies spanning the past four decades~\cite{sportperformance86, groomcoachperceptions, groomvideo}.
With the development of more advanced computer-aided systems, it became possible for coaches to manually augment the videos with annotations and markers\cite{kinovea}, then interactively play back specific sections of the videos, which was shown by O'Donahue to have a beneficial effect on training and performance~\cite{odonovideo}.

As Computer-Vision (CV) has become more sophisticated, the automation of some these video-annotations has become more accurate and prevalent~\cite{cvinsport}, but currently seems limited to semi-accurate player-location-detection, and fairly-inaccurate pose detection, which are very useful for overall analytics in some sports~\cite{pansiottenniscv}, but not for specific technique recommendations in others.

\subsubsection{Marker-Based Motion Detection}
Some systems, using multiple cameras along with reflective VICON markers placed on the body, can accurately detect movement.
This has lead to successful automation of technique-coaching in rowing~\cite{automaticrowingcoach}, yet probably has limited functionality for the average climber due to excessive setup and equipment costs.

\subsection{Wearables}
If a single acceleration-recording wearable device can capture data that is rich enough to distinguish between levels of ability and technique, that was a very useful potential feature in the early idea-development of my project.
In Ohgi's paper~\cite{oghiswim}, he showed that by attaching a small tubular device to a swimmer's wrist, and graphing the tri-axial time-series acceleration data that was recorded, it was possible to determine the full path of the arm-stroke, and distinguish between a stroke with good technique, and a later, more fatigued stroke with worse technique.

Building on this, in his review paper~\cite{callawayvideoacccomp} Callaway showed that for some sports, such as swimming, having multiple wearables on an athlete's body, providing accelerometer and gyroscopic measurements, can provide much better data and analytic feedback than the traditional video techniques.
Instead of giving an overall impression of technique, measuring the angles, accelerations and velocities of various body-parts at different phases of swimming-stroke could result in technique-adjustment recommendations, although in that paper it was noted that the relative infancy of the measurement technologies available at the time caused some accuracy issues that meant these recommendations were not precise enough for real usage.

\subsubsection{Mobiles}
The use of mobile phone devices to provide data for sports technique analysis seems to be an area with very little research.
A common use of mobile devices is through using the accelerometer to provide a pedometer functionality, counting footsteps, a feature that comes built-in with many modern smartphones.
However, this has been shown to be very inaccurate without extensive calibration, and often varies greatly between software and hardware used, with only GPS showing an accurate measurement of distance or speed travelled~\cite{pedometer}.

This lack of research into the area of mobile-usage for sports analysis could be attributed to the generally elite-athlete-orientated nature of the field;
often very specialist hardware can be used, as the recreational sportsperson is not usually the target population as it is in this study.

\section{Climbing-related Research}
Although climbing has not seen as much research interest as many other sports, there have been some studies, looking into both data-collection of athletes and a couple beginning to explore the interaction of climbers with devices and software.

\subsection{Data-Collection}

\subsubsection{Computer Vision}
Sibella et al performed an analysis of twelve climbers on a specially-constructed 3mx3m route, capturing their motion with reflective markers and a series of cameras~\cite{centreofmass}.
By using computer vision to analyse the climbers' centres-of-mass, the force, agility and power of each participant could be determined, and further testing showed that the better climbers minimised power and climbed more efficiently than the less experienced in the test group.

\subsubsection{Wearables}
Multiple studies have looked into using wearable devices to record accelerometer data and correlate that data to various performance measures.

ClimbBSN~\cite{climbbsn} uses a specially-designed ear-worn  sensor to record tri-axis acceleration data, which is then analysed with a combination of Principal Component Analysis and Guassian Mixture Models to present climbing profiles for the four test climbers.
This was used to show that the data collected from the sensors correlates strongly with the grade at which the climber was proficient at climbing, and could therefore be used to track progression in the future.

ClimbAX \cite{climbaxstudy} is a similar system that uses wearable acceleronmeter-recording bands on all four limbs, over a series of climbing sessions.
Data from many climbers across many climbs was analysed, leading to the training of a set of classifiers that could accurately assess performance, across a range of classifiers such as stability, speed, control and power.
This lead  to successful predictions of results in a later climbing competition.
The study presented itself as the first step towards an ``enthusiast-orientated coaching system", and currently a startup company of the same name have been working on building that into a cloud analysis platform, but with no product released to market yet.
It shows the potential of an accelerometer-based automatic analysis system, something that I decided to lay as an option in later wizard-of-oz studies.

\subsection{Climbing-HCI Research}
Various aspects of adding computer-interaction to climbing have been studied in recent years.

\subsubsection{Visualisation}
A prototype visualisation platform for the above ClimbAX system was developed, analysed and evaluated by Niederer et al~\cite{niederervis}.
Their logbook-style interactive web-app, which implemented and displayed the analyses developed in the ClimbAX study in a clean interface, alongside graphs of performance over time, was well-received in their usability study and reviews.
Taking from this that usability and simplicity are key, alongside the participant's positive review of graphed accelerometer data, I nevertheless wanted to perform my user-centred iterative design from a relatively clean, slate, and not allow these findings to impact my surveys and discussions with the testers I was working with.

\subsubsection{Projectors}
Kajastila et all developed a system that uses an array of projectors and laser-sensors to illuminate a climbing wall~\cite{projectedclimbwall}.
This succeeded in giving the ultimate real-time feedback with no attached sensors: climbers could climb like normal but see either games, statistics or timers projected above and around them.
However, the very specialist apparatus must be set up and calibrated for a specific wall with specific holds, meaning that although derivatives of the technology have been sold to various climbing centres, it is not feasible for the individual climber to use unless they travel to one of these few walls whilst they are set up.

\subsubsection{Setters}
One climbing-related HCI study that didn't directly interact with climbers was StrangeBeta\cite{strangebeta}, which linked mathematical chaos with machine-learning to assist route-setters - the people who design and set indoor routes.
This showed the power that augmenting normal climbing-related activities with data could create novel and more interesting routes, although some participants stated they didn't always appreciate input from the software.
It also demonstrated the complexity of the computerisation of climbing routes, requiring the development of a domain-specific-language and advanced parsers to be able to allow the app to reason about and communicate hold-variety and -location.
This intricacy (and occasional failure) for machine-learning systems to work with the very complex range of climbs, even limited to indoor routes, was a consideration that I was conscious of throughout my project.

\subsubsection{Custom Holds}
Edge~\cite{edgeinteractive} is a system involving custom climbing holds that have built-in sound, light and haptic feedback mechanisms, and link to wrist- and ankle-bands.
It guides the climber up the wall, indicating which limb should contact which part of which hold as the climber, resulting in a variety of different routes for a single route set.

\section{Other Climbing Technologies}
\subsection{Products and Apps}
\subsubsection{Logging}
The traditional form of data-collection for climbing is in the form of a logbook: a small booklet that climbers can use to log each climb they successfully ascend, with notes on date, time, location, grade, type and ease.
This has been reflected in a multitude of mobile applications and website that supply the same basic feature: to record and list climbs.
Some of these apps also have graphing features, showing the progression of the average and the maximum grades of climb summitted over time~\cite{verticallife}.

\subsubsection{Logging aided by accelerometer data}
One interactive device that was lauded by the press upon initial announcement was the Whipper~\cite{whipper}, a clip-on accelerometer linked to an app, that made over twice its funding goal on IndieGoGo but unfortunately never came to fruition.
The basic idea was a way to accentuate logbook apps with some more accurate data-collection through acclerometer, GPS and barometric (height) sensors.
They also claimed to be able to automatically determine the name and grade of the climb just from this data - a tough task that many commentators believed led to the company's eventual downfall.

Despite the rising popularity of fitness-tracking smartwatches from manufacturers such as Garmin and FitBit, these devices do not have any functionality for climbing-specific recording other than logging time spent doing the activity.
With Apple's latest smartwatch allowing app developers to access raw accelerometer data, some apps have been developed that use this to passively track a climbing session, recording number of climbs and time spent climbing vs resting~\cite{chalkprint}.

\subsubsection{Computer-aided Technique Analysis}
Lattice Training is a company set up by two climbing coaches and personal trainers, with the aim of assessing and planning climbing training with a rigorous data-led approach~\cite{lattice}.
They use a variety of strength and endurance tests on specialist equipment, and after analysing the data of hundreds of climbers, have developed a series of statistical measures that can link each strength test to expected grade of climbing ability, highlighting weaknesses to be worked on.
They have also released an accompanying mobile app for logging workouts. 
Their system has shown that by using data-collection for meaningful training over long periods of time,  climbers can progress much more efficiently and quickly, but this expensive and intense product is too much for the average recreational climber.

\subsection{Where AugKlimb fits in}
The main aim of this project was to develop a product that fills the gap between the simplistic lists of climbs presented by the logging apps and the elite-level coaching or data analytics offered to the athletes.
Something for the average recreational climber to use in a more casual every-day fashion, but that still provides meaningful data augmentations.



\chapter{Project Execution}
\label{chap:execution}

\section{Introduction}
The main activity of this project was the development of a mobile app that can usefully analyse intermediate climbers' technique and then provide them with various data whilst they climbing, whether it be for training or fun.
To do this, I performed a User-Centred iterative design process, with four large main iterations, and many smaller ones to guide along the way.

\subsection{User-Centred Iterative Design}
Throughout the execution of the development of this project, I followed the guidelines of both the Iterative Design and User-Centred Design methodologies.

\subsubsection{Iterative Design}
\begin{figure}[h]
\centering
\includegraphics[width=10cm]{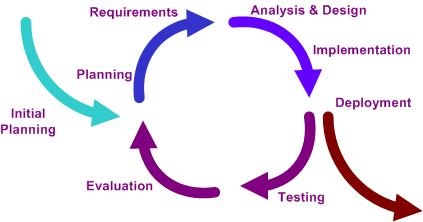}
\caption{Diagram of the Iterative Development Model. \newline \textit{\small Westerhoff [Public domain], via Wikimedia Commons}}
\label{fig:iter}
\end{figure}

Iterative Design is a methodology where a cyclic process of prototypes, tests, analysis, and refinements of a product are made.
See Figure~\ref{fig:iter}.

After some initial plans have been made, a set of requirements are laid out, then some very basic form of the product is implemented, before being tested to gain feedback, then this feedback is evaluated to inform the next cycle's planning stage.
This process then repeats, gradually refining the product and gaining stronger insights into the user's requirements.
In this project, four major iterations took place, which the majority of this chapter is dedicated to outlining.

\subsection{User-Centred Design}
Because this project aimed to build a product that most closely matches the needs and wants of local recreational climbers, I decided to also use the User-Centred design approach, which states ``users should be involved throughout the project development
process"~\cite{ISO9241-210}.
In order to do this, I met and discussed with the users throughout every stage of the Iterative Design process detailed above, not just the Testing Phase.

Therefore, within and alongside the larger iterations, minor iterative loops took place, where users were consulted repeatedly, their test-responses evaluated and suggested changes implemented, during the implementation of each major feature.

\section{Iteration \#1: Wizard-Of-Oz Prototypes}
In this section, I will describe my first large iteration, in which I gathered initial thoughts and requirements from climbers via a survey, and then trialled some of these in a series of wizard-of-oz (WOZ) tests to glean the probable requirements of the users.
Then, after implementing a prototype app, I brought it back to the users again to test, evaluating their responses to bring me forward.

\subsection{Initial Survey}
To inform my first forays into the project, I released a questionnaire online.
The ethics application for the survey (which also covered prototype testing) can be found in Appendix~\ref{appx:ethics1}, and the survey questions can be found in Appendix~\ref{appx:survey}.
I shared links to this survey across a variety of local climbing club pages, and received 49 responses in total. 
It must be mentioned that approximately half of these responses came in after I had already started development of the app, so although they did not all contribute to my initial development, I regularly checked the survey results and took into account any new opinions.
This is one benefit of an online survey: I got a much broader reach, quantity and range of responses compared to if I had performed a more traditional paper survey at a climbing wall.

\subsubsection{Survey Respondents}

\begin{figure}[h]
\centering
\includegraphics[width=10cm]{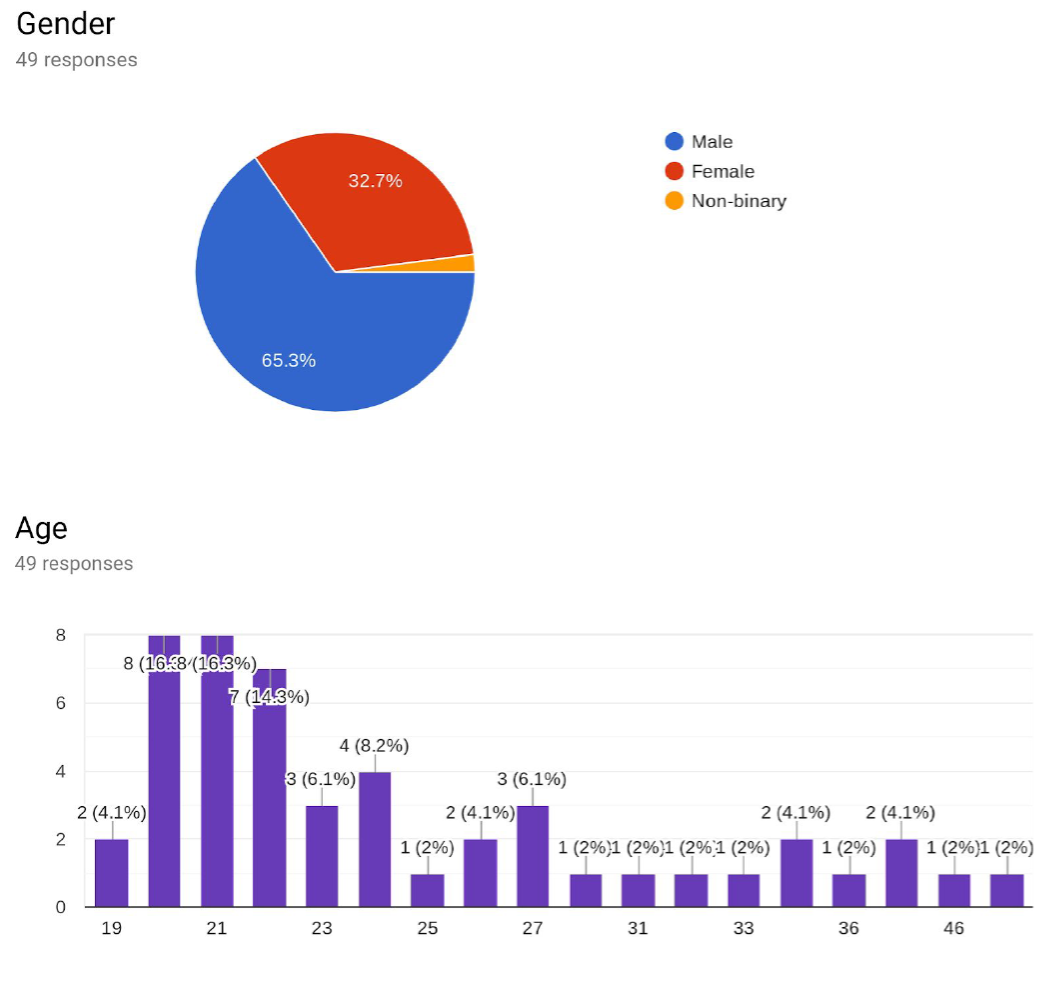}
\caption{Demographics of Respondents to the Initial Survey}
\label{fig:surveydemographics}
\end{figure}
Some statistics on the demographics of the respondents can be seen in Figure~\ref{fig:surveydemographics}.

The male-female gender split is fairly consistent with the general indoor climbing population: 65\% male respondents compared to the 64\% stated in Rapelje's study on the demographics of climbers~\cite{climbing-sub-worlds}, and 65\% aged 19-24, compared to 60\% in that study.
This was aided by efforts to spread the survey link across a range of local climbing pages and Facebook Groups, not just the student groups to get a more representative sample of views.

With three-quarters of the respondents self-describing as intermediate climbers, the survey predominately hit the target audience for the project. 
Although all the responses were taken into account, a more heavy weighting was placed on the suggestions given by intermediates compared to the lower and higher abilities.

\subsection{Insights Gained From the Survey}
\subsubsection{Equipment Used}
To inform my decisions on what form my interactive device should take, I asked what items of equipment climbers often bring to the wall.
Everyone stated that they bring their own shoes, 90\% stated they bring a chalk-bag (a small hip-bag that contains powdered chalk for drying fingers and improving grip), and 31\% said they bring a small brush for cleaning dirty holds.

This provides a range of locations for locating a potential device, either inside the pocket of the chalk-bags, or attached to a brush, keeping with the light-weight aims of the project.

Only one respondent in this question included a phone in their list of equipment brought to the wall, and one even explicitly stated that they deliberately left their phone in their bag in order to "get away" from it. 
However, in later questions, an app (and thus a phone) was the most discussed potential device for usage, with multiple suggestions that a "you would severely limit your possible audience by having the need for a device" as "everyone has a phone" so I should "stick with an app".

\subsubsection{Communication With Others Whilst Climbing}
Because a potential use of the device was to aid some of the common interactions that climbers had with friends at the wall, two of the questions in this initial survey asked what kind of things people wanted to hear whilst climbing, and what they often told their friends.

The most common response was ``beta", which is a colloquial climbing term for advice on how body-positioning should be used to solve a tricky climb. 
This can vary a lot between climbers, with each climb usually having two or more potential solutions.
Also, being able to ``read" a climb to determine this beta is a tough and much-sought-after skill within the climbing community, to the extent that it is often nearly impossible for the very best in the sport, and thus definitely out of reach of any Artificial Intelligence (AI) I could hope to create within this project.
An alternative, which was suggested by a few survey respondents, could have been to record (with either video or other sensors) a proficient climber climbing that route, and then showing that at a later date. 
However, this data-collection and -replay contradicted my initial aim of the device/app being able to be used anywhere by anyone: for a low-cost app to be usable everywhere, the requirement to go to every climbing centre every few months and re-record somebody climbing every climb at each climbing wall, just for the product to remain usable, was not feasible.

Another common interaction highlighted by the survey was ``pointing out holds" to a climber, who is often so engrossed in the climb that they do not notice a certain location that they could put their hands or feet.
This has an obvious potential use-case for a device, by using a colour video and existing computer-vision algorithms for coloured-blob-detection to find and then audio-relay the location of nearby holds to a climber, so this was selected to be examined further.

\subsubsection{Requested Potential Features}
Arguably the most useful question was the one asking
\textit{``Are there any features you'd like to see in a training app or device?"}.
This elicited a very wide range of suggestions, from yoga and meal plans to VR-viewing of someone climbing alongside.
Many of these were either too simple (and could be achieved either by existing apps, or an online search), or far too ambitious in scope.

The most common request was a way to log climbs in some way, and although a few logging apps already exist, this feature was definitely required in some way alongside whichever novel features were to be developed.
Tracking progression and seeing gradual improvements over time is a typical characteristic of training in any sport, but instead of just listing climbs that had been done, a few respondents suggested that other metrics could be tracked, such as height climbed, frequent muscle groups used, and weak-points in technique that caused a climber to fall off at a certain point.

Many respondents emphasised that ease-of use was very important, so they didn't waste time climbing by clicking through the app. ``Large buttons for chalky fingers" were requested, alongside ``a minimum of two clicks between opening the app and using it".

The video-analysis of technique was mentioned by seven respondents, with a mixture of technique-tutoring, centre-of-gravity detection and limb-annotation suggested.
Both centre-of-gravity and limb-annotation could be feasible achieved with computer-vision, so those were taken forward to the WOZ testing phase.
However, if beta-detection has been ruled out as too difficult for CV/AI, then giving accurate technique advice is even further out of the range of potential features: with the high complexity of movements in the sport, years of learning required to achieve coaching qualifications, and the high risk of injury if incorrect, giving explicit advice about how to climb is not something that I wanted to go into with this project.

One metric that was proposed is the analysis of efficiency or ``smoothness" in a climb.
A commonly accepted signifier of ``good technique" is when a climber moves neatly and directly between each successive climb, without any jerky motions or unnecessary readjustments~\cite{centreofmass}.
An accelerometer recording could be used to determine this metric and thus this suggestion was also brought forward to the next stage of testing.

\subsection{Feasibility and Potential Limitation of Video Analysis}
The initial survey showed interest in using CV to either give feedback about climbing technique, or to point out where nearby handholds were.
To those ends, I began experimenting with using OpenCV to detect blobs in previous recordings of my own climbing.

Although some small success was had in detecting the centre of body mass, and some nearby coloured handholds, the wide variety of different lighting conditions and noise from the low-res camera made the detections very inaccurate.
Another issue that arose was that the code would analyse the video very slowly, even on my PC with a graphics card, so converting that to battery-efficient code that could run in real-time on a mobile device would have been extremely difficult.
Due to the time-scope of this project, and the aims being to iteratively develop something and analyse how climbers interacted with such a product, I was wary of spending too much time trying to get video-detection working.

\subsection{Wizard-Of-Oz Tests}
To determine the requirement for this video feature vs using an accelerometer to provide data input, a series of informal discussions and WOZ tests were performed at a climbing wall with 3 boulderers.
This involved me writing a script, and showing images of potential screen displays when they performed an action or climbed.

Now knowing the probable abilities of the CV's video-analysis, which was either to label limbs and centre of mass on a video output, or to state nearby handholds during the climb, I acted out both of those features in a WOZ format, and the results of that test was that they were determined to be potentially useful.
However, the labelled video feature was stated to be only slightly better than just watching a video playback, and the inefficiency of setting up a phone to point at a wall to call out handholds with some delay was not something that excited the climbers.

The WOZ prototype that involved having an accelerometer data showing up on a graph with some form of "smoothness rating" was received well in the testing, and I felt that it would be a good idea to explore that route with the iterative testing and design, whilst also working on some of the video analysis in the background.

\section{Iteration \#2: Adding Accelerometer Data}
This section will outline my second large iteration.
Now that I had completed the first iteration, with WOZ prototypes, and roughly knew which features and form-factors the users wanted, I next had to design and implement a real-code prototype to test.
In order to do this, I had to make some choices regarding the tools I'd be using, then make a minimal app that could record some data, before exploring different output options with my testers.

\subsection{Fundamental Early Decisions for Development}

\subsubsection{Platform Choice}
Despite researching a variety of different devices and form factors, one of the key aims of the project was always to make the final product as accessible as possible, therefore a OS-independent mobile app was decided upon after reviewing the survey results.
By not using any extra equipment such as wristbands or 3d-cameras, the scope and ability of the final product could be limited (by lack of sensors and processing power).
However, limiting the need for anything but just a phone, which almost everyone owns, was a worthwhile sacrifice.
As well as making user-testing much easier, this choice also helped narrow down the direction of the project: working solely within the limits of what a mobile app could achieve meant that the capabilities of that platform could be fully stretched and explored.

The two inputs that mobiles can easily capture, and that would potentially be most useful for the analysis of climbing technique, were video recordings and accelerometer data.
Video recordings can be analysed with various Computer Vision (CV) techniques and accelerometer data can be shown on a graph and analysed statistically to provide various outputs.

\subsubsection{Tool Selection}
\paragraph{App-development tool}
Next was to find a app-development tool that is quick and easy to use (for quick repeated iterations of the app), can easily import or link to the OpenCV Library (for the CV aspect), and is platform-independent (so any mobile phone owners can use the app, irrespective of OS).

Unity was chosen as it meets all three of those requirements, with a wide variety of ``Assets" (downloadable open-source libraries) that can be imported.
 
By providing a platform for creating simple user interfaces that could be exported to a range of platforms, and easy linking between the UI elements and back-end \verb|C#| code, I knew that quickly building and iterating over multiple app designs would be possible.
Also, having used Unity for games development in the past I was very familiar with it, so compared to potentially wasting time learning how to use other tools, this one met all the above requirements, with the added benefit of previous experience using the tool.

\paragraph{Version Control}
For version-control I used \verb|git|, backed-up on GitHub.
Being a one-person project I rarely bothered with the overhead of using different branches, but the ability to roll-back to working code and releases, and to \verb|stash| and \verb|pop| various files at different times was very useful during development.
Once or twice a week, after a new feature or big set of fixes had been added, I would up-version the app, and release a compiled .apk binary to the project's GitHub page.
This allowed the easy re-installing of older versions when trying to locate a bug, as well as a clear documentation of all fixes and features included at each minor version upgrade.

\subsection{First minimalistic app: Recorder}
After the success of the accelerometer idea in the previous iteration's tests, I decided to build an app based around this, to use in some real in-field testing.

This first app just had a two buttons, START and STOP, which would, respectively, start and stop recording the mobile phone's raw accelerometer data into a \verb|.csv| file.

\subsubsection{Axes Selection}
The accelerometer's API allowed me to collect data from each of the individual axes, so it seemed like a good idea to use this to determine and compare acceleration vertically vs horizontally, which was a potentially useful feature.
However, after viewing the raw data output from the first climbing session, it became clear that the phone's axes would very rarely match up with the real-world axes in any meaningful way, and even some sort of calibration at the start of the climb would quickly become meaningless as the climber rotates and moves up the wall.

Therefore just the total magnitude of the vector-sum of the acceleration measured along all three axes was used going forward.

\subsubsection{Mock Output}
\begin{figure}[h]
\centering
\includegraphics[width=12cm]{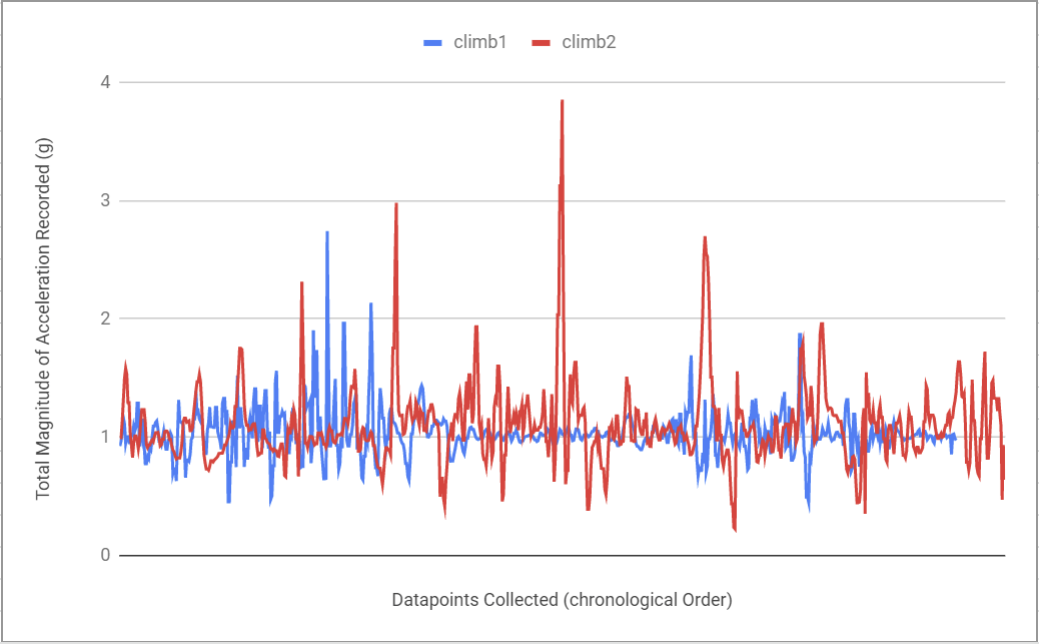}
\caption{Overlaid Graphs of the Accelerometer Data Recorded During Two Different Climbs}
\label{fig:twoclimbsgraph}
\end{figure}

During this first test, the lack of an in-app output was an obvious limitation, but after a climbing test session with just this feature, I was able to copy the data into a spreadsheet software and graph the results, giving Figure~\ref{fig:twoclimbsgraph}.

I then met up again with the climbers from that test-day at the wall, and we discussed the graphs and could recognise which climbers were climbing which climbs, just from the spikes on each line graph; this was the first time I saw a clear correlation between the acceleration and the climbing style and ability, an exciting result.

\subsection{In-App Statistics}
The testers enjoyed viewing the graphs I had produced, so I started working on adding graphs to the app.
Unfortunately here I hit the first limitation of the tool I was using - Unity.
Primarily designed as a Games Engine, Unity didn't have an easy method of plotting graphs, and any Asset Store solutions were expensive to buy. 
A solution involving drawing dots on the screen for every data point, and calculating the size and angles of lines connecting them was possible, but I knew that would take a few days of coding, and wanted to iron out any potential kinks in my data-collection methods before investing that much time into coding.
Also, I had another testing day scheduled with my climbers, so wanted to quickly code a simpler form of output to see how they interacted with the data.
Simply calculating the min and max acceleration, and the time taken to climb, only took a few lines of code, and so the second prototype of my app included a screen that would display those statistics after a climb.

During the test session, both of the people I was climbing with enjoyed using their max acceleration in a fun and competitive manner.

They took it in turns trying to produce as much acceleration, or "power", as possible on some climbing moves, and also compared times whilst slowly doing easier climbs, trying to keep their max acceleration down to as low a score as possible, and doing certain climbing moves with as much or as little speed as possible.
When the potential of a "smoothness" score was brought up in discussions, they told me that a statistic other than just max acceleration, and more closely linked to actual climbing technique, would be very useful.

\subsection{Development of Smoothness Score}
Seeing a range of graphs on the spreadsheet software (similar to Figure~\ref{fig:twoclimbsgraph} but with many more climbs) was very interesting.
Some climbs' graphs were consistently fluid with no big spikes, some were mostly flat with a few large spikes, and some had a lot of small spikes, which clearly showed that there were discernible differences between the accelerometer data produced by different climbs, something I intended to utilise.

After discussion with some climbing coaches, they suggested that what they often tell clients to do is to climb the same easy climb repeatedly, aiming for as fluid and smooth a motion as possible.


Alongside this, occasionally some harder climbs call for bigger ``dynamic" moves, but accuracy in catching the holds without lots of small adjustments is still important for good technique.

What the coaches I spoke to said they would generally judge as a climb with poor technique is one where many small inefficient and jerky movements were made: the repeated re-gripping of holds can lead to early forearm tiring, and the uncontrolled and ungainly fluctuation of momentum of the overall body mass is very inefficient.
One aim of good climbing technique is to prevent this early tiring, and reduce as much inefficiency in movement as possible, as these are often the factors that prevents the successful summit of a harder climb.

However, because climbing technique and style can be very variant among both climbers and between each climb on a wall, there was no way to provide an accurate "best" or "aim" score, especially when variations in the acclerometer recordings on different phones is taken into account.
Instead I opted to try and find some form of quantifiable metric that, as the testers had suggested, was more closely linked to climbing style than just minimum or maximum acceleration, yet didn't have an absolute limit or goal, to keep the flexibility required between different climbs and different sensors.

\subsubsection{Four Alternatives}
Although previous studies\cite{climbaxstudy, climbbsn} use machine-learning techniques to train classifiers that can predict ability, attempting to directly analyse the smoothness of a graphed acceleration data-set to distinguish between climbing styles has not previously been attempted.

I decided to test out a range of different options, on real climbing data, to determine which score was the closest representation of climbing style.
To aid with the comparison of climbs, the metric should be able to distinguish between different styles: if a climber is repeatedly trying to climb as statically and smoothly as possible, then the score should be able to reflect that by changing in a predictable manner.

\paragraph{Mean}
$$\frac{\sum n }{N}$$
The simplest statistical measure was worth testing: a large amount of higher accelerations would result in a higher mean, which would possibly show differences in climbing style.

\paragraph{Variance}
$$\frac{\sum (n - \bar{n} )^2 }{N-1}$$
Again, a classical statistical metric that had potential: measuring how varied the accelerometer data-points are is linked to the above discussion on how more ``spiky" graphs often portray less smooth climbs.

\paragraph{Average squared difference between each consecutive data-point}
$$\frac{\sum (n_i - n_{i+1} )^2 }{N}$$
Taking variance a step further, and calculating the pairwise differences in the dataset could give a good measure of how ``jumpy" the chart is.
By measuring the differences in acceleration, this statistic works in much the same way as finding the average squared ``jerk", which is the derivative of the acceleration.

\paragraph{Single-Lagged Auto-Correlation}
$$
\frac{\sum(n_{i} - \bar{n})(n_{i+k} - \bar{n})}
      {\sum(n_{i} - \bar{n})^{2} }
$$
My idea for using this metric was that if each data-point is closely correlated with its successive data-point, then the overall line would have less jolts, but run in a more smoothly varying way.

\subsubsection {Comparative Testing}
I had three test climbers climb three different routes, in either a deliberately static style, a deliberately dynamic style, or their usual style: a hybrid of the two.
Using the data recorded from these climbs, I calculated each of the above metrics, visible in Table~\ref{tab:smoothnesses}.

\begin{table}[h]
\centering
\begin{tabular}{|c|c|c|c|c|c|c|c|c|}
\hline
climb  &  climber & style & min & max & mean & var & var-diff & lagged-autocorr \\ \hline
purple v0 route & 1 & static  & 0.29 & 3.50 & 1.10 & 0.12 & 0.06 & 1617.90 \\ \hline
purple v0 route & 3 & static  & 0.07 & 6.39 & 1.09 & 0.16 & 0.12 & 1600.46 \\ \hline
purple v0 route & 2 & dynamic & 0.16 & 10.96 & 1.11 & 0.39 & 0.18 & 2150.63 \\ \hline
purple v0 route & 3 & dynamic & 0.20 & 9.76 & 1.14 & 0.40 & 0.16 & 1931.72 \\ \hline
yellow v1 route & 1 & static  & 0.21 & 2.20 & 1.06 & 0.05 & 0.02 & 795.96 \\ \hline
yellow v1 route & 1 & hybrid & 0.06 & 11.41 & 1.06 & 0.17 & 0.11 & 2176.12 \\ \hline
yellow v1 route & 1 & dynamic & 0.03 & 8.20 & 1.07 & 0.21 & 0.10 & 1862.94 \\ \hline
pink v2 route & 2 & static & 0.45 & 2.74 & 1.04 & 0.04 & 0.03 & 793.59 \\ \hline
pink v2 route & 2 & hybrid & 0.21 & 4.04 & 1.05 & 0.08 & 0.03 & 1939.77 \\ \hline
pink v2 route & 2 & dynamic & 0.16 & 10.32 & 1.12 & 0.25 & 0.15 & 1826.36 \\ \hline

\end{tabular}
\caption{Comparison of Smoothness Score Candidates}
\label{tab:smoothnesses}
\end{table}

As can be seen, despite both the var-diff and the lagged-autocorr algorithms showing some promise, the only metric that can reliably distinguish between the three styles of climbing is the variance, so this is the score I took forward.

\subsubsection{Confirmatory Testing of Variance}
In the app, I added a calculation at the end of the accelerometer-recording, which would display the variance to the user on the app screen immediately upon completing a climb.

Taking this back to the test-users, they agreed that this metric was much better for them being able to compete on a climb, both with themselves and with each other, to get "as static a climb as possible", meeting the requirements for this score.

One point they did raise was that comparing a score of, for example, $0.21234$ and $0.17018$, felt odd.
My testers found comparing and discussing numbers in the 20-100 range much easier and more natural, so on the next iteration of the app, before displaying the variance I multiplied it by a hundred and displayed it as an integer.

\subsection{Graphing Acceleration Data}
Now that I had a working "score" that I could analyse user interaction with, I moved onto the next feature that my testers requested: plotting and showing a graph of the accelerometer data in the app.

\subsubsection{Timing Issues}
For the first version of my graphs, I simply plotted a rudimentary scatter-graph:
within a given box on the screen, for each data-point $n$ in the accelerometer recording, I drew a small circle object $n$ pixels up, and spread the dots across so they stretched to fill the width of the box.

Although this resulted in what looked like a line-chart with time as the x-axis, there were a few issues that arose when I took this version of the app to be tested;
the graphs that were being displayed didn't quite ``look right".
Although the x-axis of the graph was supposed to represent time, because I wasn't recording the actual time but just successive data-points, the x-axis didn't truly represent time in the real-life stress-testing that climbing involved.

What I found out to be happening was that the way Unity handles incoming accelerometer data was to record it as fast as possible - which is exactly what a developer would want if they were using it as an input to a game.
However, this caused the side-effect of the time-scale varying slightly, in a usually unnoticeable manner.
During at-home debugging, when I recorded a timestamp to the csv file alongside the accelerometer data, the time differences only varied from 0.004s to 0.006s, and although this would obviously need to be fixed for the accuracy of the graphs, it didn't explain how the graphs when climbing would be so inaccurate.

Taking this code back to the the climbing wall a few days later, I discovered that during some more dynamic climbing moves, the differences in recordings reached up to 0.5s.
Upon closer inspection, I saw that when the phone was detecting a very large acceleration along with a large rotation, it would try to change the display orientation, resulting in a brief pause in the recording of the accelerometer.

Therefore, to fix these issues, I forced the app to remain in portrait mode, and re-wrote the recording function caller to ensure that exactly 20 samples per second would always be recorded.

Despite fixing this consistency of timings, I also re-wrote the graph-drawing code to take into account the associated timestamp for each point.





\subsection{Viewing Previous Climbs}
Up until this point, the statistics and graphs of each climb had been only visible directly after a climb had been completed, and to compare climbs the users had to memorise the previous climb's smoothness score and roughly how the graph looked, to then hope to improve.

Viewing a list of all the previous climbs, with associated smoothness scores and graphs, was the obvious (and much-requested) next feature to add.

\begin{figure}[h]
\centering
\includegraphics[width=6cm]{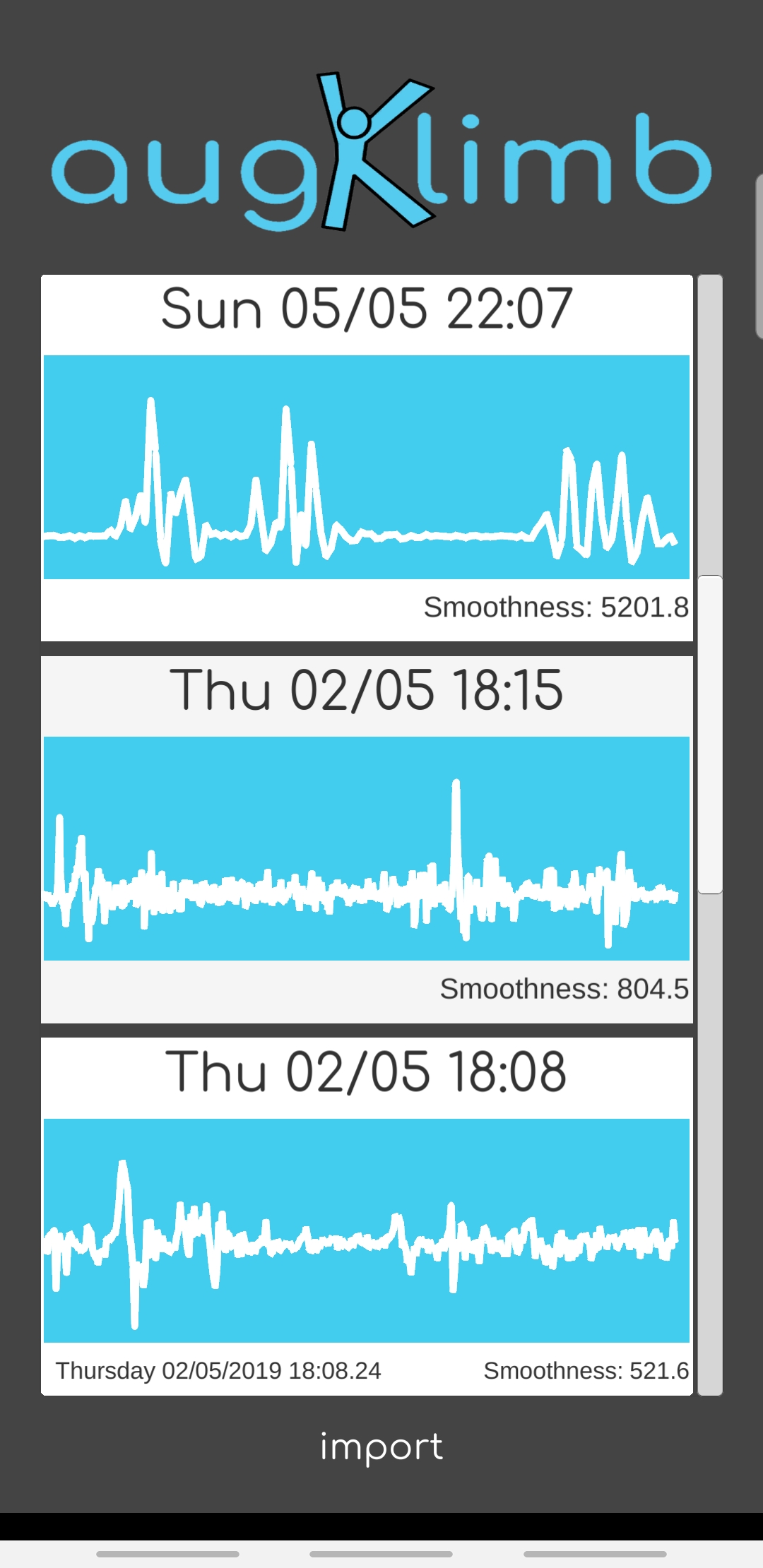}
\caption{augKlimb's scrolling view-all-climbs page}
\label{fig:scrollview}
\end{figure}

Because of the way that I had been saving the accelerometer recordings to the phone's persistent memory for debug purposes, I could scrape that folder for all csv files, and populate a scrolling list with a set of graphs and smoothness scores.
See Figure~\ref{fig:scrollview} for the final display of this.

\subsection{Viewing Individual Climbs}
\begin{figure}[h]
\centering
\includegraphics[width=6cm]{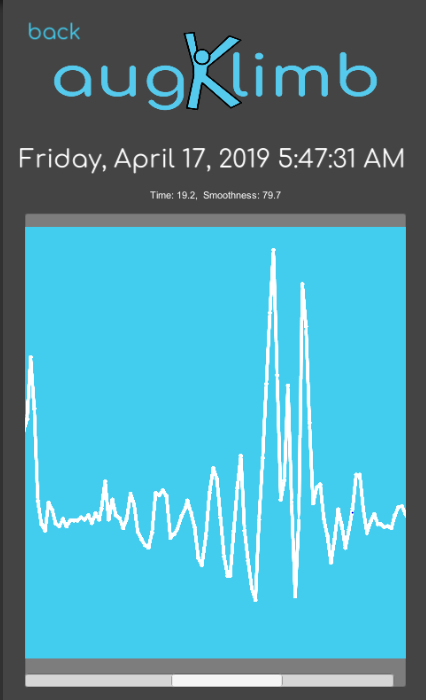}
\caption{augKlimb's single-climb-viewing page}
\label{fig:singleclimb}
\end{figure}

By viewing how my testers used this scrolling view, to compare previous climbs' graphs and smoothness scores, I also noticed that they would occasionally touch a climb's graph with a finger.
When asked about this action, they told me that they had instinctively clicked, with the expectation that they could load that specific climb and view more information on it.
Therefore I added another page into the project (see Figure~\ref{fig:singleclimb}), to view an individual climb's graph in full-screen detail, with more statistics like smoothness and time taken all on that screen.

\subsubsection{Horizontal Scale}
With the full-screen view of an individual climbing graph now possible, a discussion of scale was needed. 
Whereas in the scroll-view of all the climbs, the graphs were scaled to fit the whole data-set inside the given box (again see Figure~\ref{fig:scrollview}), one of the main expectations of the per-climb graph view was that graphs would be "zoomed-in" to see more detail.
After some testing it was found that a scale of 100px per second was optimum for viewing the acceleration's spikes with enough clarity whilst also not making the longer climbs to have an excessively long scroll to view. 

\section{Iteration \#3: Adding Video Feedback}
In this section I outline the third major iteration cycle: the incorporation of video footage and frame-by-frame playback linked to the acceleration graphs created previously.

One tester had suggested that as it was now possible to scroll horizontally through a climb's graph, using that scrollbar to also view frames of a connected video file could be a useful feature.

One option would be to use the phone to record a video, and get the accelerometer data from a wristband or other device.
However, keeping in with the theme of accessibility and only using a mobile app, the idea of connecting two mobile phone devices, so a friend with the app could video-record you as you climbed with the accelerometer feature, was chosen.

\subsection{Networking Options}
\label{network}
A variety of different options for networking two phones together was explored, including Unity's built-in Networking, Wifi-Direct, uploading to the Cloud, and Bluetooth.

These options had varying levels of support through Assets on Unity's Asset store, but after looking through reviews most of them were deprecated, broken, or did not fit my needs.
Those that potentially did cost upwards of \pounds50, which is not something I was willing to pay for a library of code that may not do what I needed.

Thus, I would have to code my own solution.

\subsubsection{Unity's Built-in Networking}
The only one of the above options that is provided in some way by Unity is the Networking libraries.
At first its probable ease of use seemed like a good option, and despite having some issues using it in the past (during my third-year Games Project)I knew how to connect various devices together.
However, the UNet library provided by Unity had just recently been deprecated, and the replacement networking service was under development at the time of writing this, so documentation was often unavailable or incorrect.

Also, despite the ability to link two phones whilst I was at home, when at the climbing walls this presented an issue as the free WIFI provided at these wall often blocked LAN connections between devices.

\subsubsection{Cloud}
Uploading to cloud would require coding some sort of scalable server and storage, something I could definitely do, but this level of overhead for what should just be the sharing of files and syncing between two devices that are located physically very close together felt like an unnecessary amount of work and time spent.
Also, it would have been very slow to upload and download a video over the web for every climb, whereas a solution that could utilise the close physical locations of the devices to transfer files faster.

The requirement to add some sort of user profiles, or unique code to differentiate and request the correct download, was a big added complexity, yet one that could have given an added bonus of users' data being backed up to the web and usable in other ways, for example from a web platform.
Although this may be something to look into in the future a cloud server isn't the solution to this specific problem, and is not really within the scope of this project.

\subsubsection{Wifi-Direct}
Wifi-Direct enables the direct connection and file-transfer between compatible devices
Although this seems like a great technology, and could be ideal for this specific use case, the few docs I could find online that detailed using Wifi-Direct with Unity stated that many parts of this feature had been deprecated, and was closely linked to the now-obsolete Unity Networking API.

\subsubsection{Bluetooth \& NFC}
An idealised use-case for the app that presented itself in WOZ testing was to just bump two mobile devices together, ``pairing" them using NFC, and then using the longer-range Bluetooth to transfer a sync command and any video or data files.

Unfortunately, this uncovered a large issue with using Unity as my tool of choice - no support or libraries for either Bluetooth or NFC. 
Even the single free Asset online was broken, leaving me with one solution: to write my own \verb|Java| plugin using Android Studio to connect the NFC and handle Bluetooth file transfers, then expose public functions in a compiled \verb|.jar| file that the \verb|C#| Unity code could then connect to.

After spending a few days writing this code, it worked occasionally, but the interplay between the two sets of libraries, on a variety of Android devices, was just too unstable.

\subsubsection{Native Messaging}
Both Android and IOS provide easy APIs for sharing files via social media or email, which is utilised by the free NativeShare Asset I downloaded from the Unity Asset Store.
Using this library, the JSON representation of each climb could easily be shared between devices using a messaging app, which was enough of a solution to this problem to allow me to take it to the testers and determine whether a more streamlined solution was needed.

During testing, climbers could record a climb, view it, and then hit the Share button to send that file to a friend, who could import that file and view it on their phone.
Despite the time spent on unsuccessfully trying to get Bluetooth working with my plugin, by simply using this same button, my testers could use Bluetooth instead of the messaging app to send the climb files - a slightly more convoluted approach to simply bumping two phones together, but one that worked more reliably when Bluetooth stopped working, which it did occasionally on one of my test devices.

\subsection{Video Player}
Now that there was a way of sharing climb-data-files, it was time to add the video-scrolling feature.

By changing the climbing graph view to shrink down to half the page, and adding one of unity's VideoPlayer components in the top half of the screen, the video could play.
Then, to set the frame of the video to display, the graph's scrollbar's position could be queried, and translated into a frame-number, which was set on the player.

Everything seemed to work great, and the testers really enjoyed it:
I would record them climbing with their phone in their pocket, then send them the video via messenger, they would download and attach it, and then could view and scroll through the video frame-by-frame to see which spikes in the graphs corresponded to which moves.

\subsubsection{Timing Issues}
However, the two recordings (video and accelerometer) would have to be started at precisely the same time in order for the timings to match up.
There was, at first, some potential in scraping the video-file's metadata for a timestamp, and matching that timestamp with the times in the recording; unfortunately when the videos are sent via the various messaging apps we were using, they would be transcode the video - to save data whilst sending, I assume - which would strip all useful metadata and timestamps.

Therefore, instead of automatically time-matching a list of recorded videos with a list of climbs, a video file would have to be manually selected and added to a climb to associate the two - an unwanted but necessary extra step for reliability.
Also, a manual ``Video Offset" selector was added above the video-player, so users could adjust the time-differential between the graph and video so that they could scroll in time if the automatic metadata-scraping and timing-detection failed to predict the correct time difference upon manual matching.
To aid this automated system, any video recordings made via the app would save the exact timestamp (to millisecond-precision) into the video's filename, which is not always changed when these messaging services transcode the video.

\subsubsection{Video Player Bugs}
Through this testing of a variety of video-sending methods, I discovered a bug:
on some devices, some videos, in some orientations, with some video-formats, would cause the video player to fail.
This is one of the toughest bugs I have ever had to fix, as it was very nearly un-reproducible: If I changed any one of the four requirements above, the bug would disappear, and if I changed the scaling settings in the \verb|VideoPlayer| component then a different combination of the settings would cause it to fail instead.

Most often, the use-case that produced this bug was when a video recorded on a phone was matched with an externally-imported climb-file.
Unfortunately, that was a very common use-case for my testers.
Whether the video was recorded through the in-app video recorder, or the phone's native camera app, it would usually fail.
If that video was sent via a messaging app, then re-downloaded off that app, it would work.
If that video was rotated and then rotated back again, it would work.

This was frustrating, as surely if a video was recorded on a device, and playable by that device, Unity should have access to the codecs required to correctly play that video?
In total, around two week's worth of development time was spent just trying to figure out how to play video reliably, as it would be very frustrating for the testers when they'd get down off a climb expecting to analyse the graphs and visualise the climb frame-by-frame, only for a flashing screen to greet them.

Eventually I completely changed how the videos were being played in Unity to solve this issue: instead of using a simple VideoPlayer component, I attached a second camera to the VideoPlayer API, and then scaled and rendered that camera's view onto a mesh in the view.
This more convoluted way of displaying video wasn't as smooth to playback the video, and delayed when searching for a specific frame, especially on older devices, but the decision was that a flickering video feed was better than an unreliable one.


\subsection{Testing}
The testing phase of this major iteration was successful: users enjoyed viewing the video footage alongside the graphs, and made a few other suggestions for UI improvements for the next stage too, but the app was nearly ready for deployment.

\section{Iteration \#4: Final Improvements and Deployment}
Now that the app was relatively stable and the two big features (acceleration and video) were working well together, it was time to clean the code with a refactor, implement the list of small requests that my testers had been suggesting, and then deploy the app, publishing it on the Play Store.

\subsection{Code Refactor}
Because I was starting to send climbing files between devices, I needed a more reliable way of exporting and parsing those files, especially with the added complexity of having video-paths and other information attached to them, the plain \verb|csv| files were not sufficient.
Also, having gradually built up the app with a script file per-view, for example one script to handle the recording and another to handle the viewing, as I added more features and file information, the code-base was getting convoluted.

\subsubsection{Refactor}
Therefore, I completely refactored all the code, keeping the only the minimum logic required for each view in those files, and splitting out all file-handling and graph-drawing methods into separate classes, along with creating a new \verb|ClimbInfo| dataclass to store the accelerometer data, title and associated video information.

Now that all the data associated with a climb was contained in a single class, I could utilise Unity's built-in \verb|Serialization|, which could automatically convert a correctly setup class directly into a \verb|json| file, and visa-versa. 
This was complex to achieve, but it proved very useful later on when adding more information to my ClimbData class: I no longer had to maintain my own file-saving and file-parsing code, but could rely on the serialisation to work with the files.

\subsubsection{Caching}
This refactor also gave me the chance to fix some performance issues that some of my testers had been having after using the app for multiple sessions.
As the number of climbs increased, so did the time it took to populate the list of climbs, as each file had to be loaded, and each graph drawn, before that page of the app could load.
Caching a list of \verb|ClimbData| objects prevented this repeated file-loading, but caused some issues with cache invalidation:
\begin{itemize}
    \item When loading the app for the first time, all of the climb-files in persistent storage should be read and cached.
    \item When a new climb is recorded, it should be saved to persistent storage and also cached.
    \item When an external climb-file is imported, it is first loaded into memory, then should be saved both to persistent storage.
    \item When a climb file is edited, it should be overwritten in persistent storage, but is already cached so should not be re-added to prevent duplicates.
\end{itemize}
This all took a lot of working out to cover all edge-cases, but the eventual solution involved tightly coupling cache-updating with the action of saving to disk, and decoupling it from any loading functions, checking for duplicates before adding to the cache, and adding an auto-initialisation feature to the cache:
If the cache is empty when queried, it will contact the \verb|FileHandler| and populate itself with a list of \verb|ClimbData| objects corresponding to what is in persistent storage.


\subsection{Testing-driven UI Improvements}
Alongside the video feature being developed and tested, my twice- or thrice-weekly testing sessions had provoke a range of smaller UI tweak and improvement suggestions, so I implemented all of those as I prepared the app for the first full deployment:

\subsubsection{Countdown}
One of the earliest of these was that starting recording as soon as the START button was pressed would record the acceleration caused by the climber putting their phone in their pocket and walking up to the wall.
To prevent this, a five-second countdown timer was introduced before recording commences, accompanied by a vibration so that the climber could have time to set up and know when to begin climbing.

\subsubsection{Guides}
Also, a lot of questions were asked when climbers first started using the app, so alongside a guide and FAQ on the website, I added clearer button names and brief descriptions at the bottom of each page to describe what happens with each feature.

\subsubsection{Cropping}
Sometimes when reaching the top of a climb, testers would just jump off, and land on the floor.
This created a large spike on the accelerations graph, and had a resultantly large impact on the smoothness score.
Thus a button was added to "crop" a climb's graph.
The horizontal location of the scrollbar was detected, and after a confirmation from the user, any climbing recording data after that point would be removed.

\subsubsection{Deleting}
To keep down clutter, testers had previously requested the ability to delete old or mis-recorded climbs, so now that each climb had its own page, a button was added that would remove the currently-open climb from the device.

\subsubsection{Back Button}
For Android, I removed the in-app back button, instead using the native OS's API to detect when the phone's back button had been pressed and returning the user to the previous screen, integrating the users' usage more closely to what they are used to from other apps.

\subsection{Publishing to Google Play}
As I don't have either a Mac or an iPhone, despite my choices to use a platform-independent software tool, I was only able to build to Android throughout this project.
That meant that the natural distribution service to use was Google Play.

Publishing to that store required a developer account, images of the app in use, various review procedures, and for me to write both a code of conduct and a privacy policy.

This Play Store listing can be seen \href{https://play.google.com/store/apps/details?id=com.lukestorry.augKlimb}{here}, with a screenshot in Figure~\ref{fig:playstore}.

\begin{figure}[h]
\centering
\includegraphics[width=9cm]{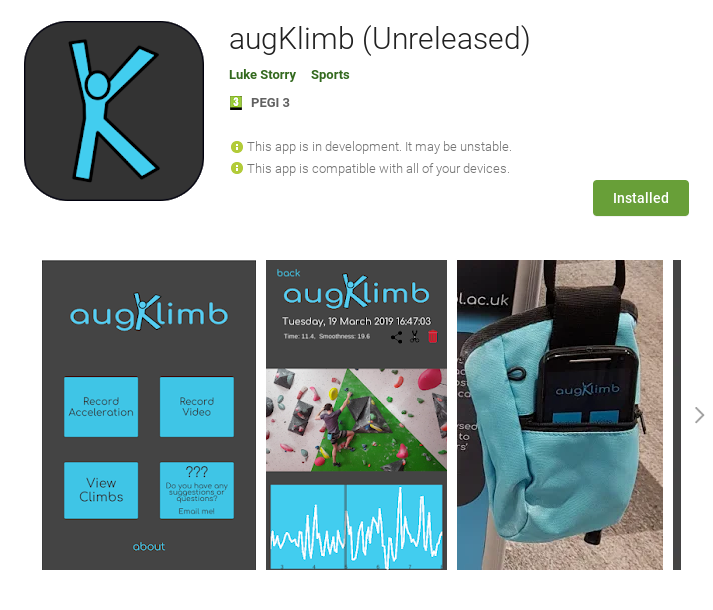}
\caption{Screenshot of Play Store Listing}
\label{fig:playstore}
\end{figure}

To ensure that the general public did not have access to the test versions of this app, but only participants who had consented to be a part of this study, I did not release a full production version of the app, but used Google's Limited Beta Rollout feature to only provide beta access to the testers that I had given links to.

\subsection{Easier testing}
One obstacle that all my testers had faced throughout the project was how to install the app.
With such an iterative development processes, I was recording the app every day and trying two or three different versions each week.
This required either me lending one of my mobile devices to the testers to use, or side-loading an \verb|.apk| file onto the users' phones, disabling various security measures to install the app each time.

Having the app on Google Play not only made it a lot easier to upgrade the app on everyone's phones before each testing session, but the added legitimacy provided by not having to side-load the apk meant that a wider variety of people felt comfortable installing the app from a provided link.
This caused the amount of testers who were regularly using and testing the app to increase from 4 to 15, providing a rich variety of feedback from regular users of all abilities, both through occasional conversations and through the feedback form I had linked to the homepage of the app.

\subsection{Smaller Test-led Refinements}
Having more testers, each using the app slightly differently, led to a wide range of feedback, advice, and suggested improvements.
Also, the testing being on an assortment of different devices highlighted a range of small bugs that were either non-existent or at least infrequent on my own devices.
Few of these are worth mentioning, usually just requiring corrections to button-placement or changing the configuration of the file browser.
One bug caused a permissions error on Nexus devices, but after a week of helping that user manually re-enable the correct permissions, Unity released an update that fixed what turned out to be a bug in their engine.

\subsubsection{Title Editing}
Until this point, the title of a climb had just been the date and time that it was recorded.
In the initial survey, a climb-logging app was requested, and so to meet that demand I added the ability to edit this title to name the climb's grade, colour, and location.

Some testers didn't use this feature as they weren't interested in creating a list of climbs, just analysing the specific one they were working on at the time, but for others this added a nice way to be able to scroll back through climbs from a previous session and easily find one to replicate or compare to.

\subsubsection{Video Recorder}
Despite previously having moved away from recording videos within the app, and instead just using the phone's normal camera app, some testers suggested that it would feel more integrated to have all of the recording features within the app.
To this end, I updated the in-app camera feature to automatically save videos both to their phone's external camera-roll and to the app's persistent storage.

\subsubsection{Colour-Grading the Graphs}
One of the most common user-flows that I observed during the testing was repeatedly climbing the same climb to try and get the lowest smoothness-score possible, then during a rest-break, scrolling through and reviewing the graph for each climb, to see which moves caused the largest spikes, and thus how they can climb in a different way to reduce the score further.

\begin{figure}[h]
\centering
\includegraphics[width=6cm]{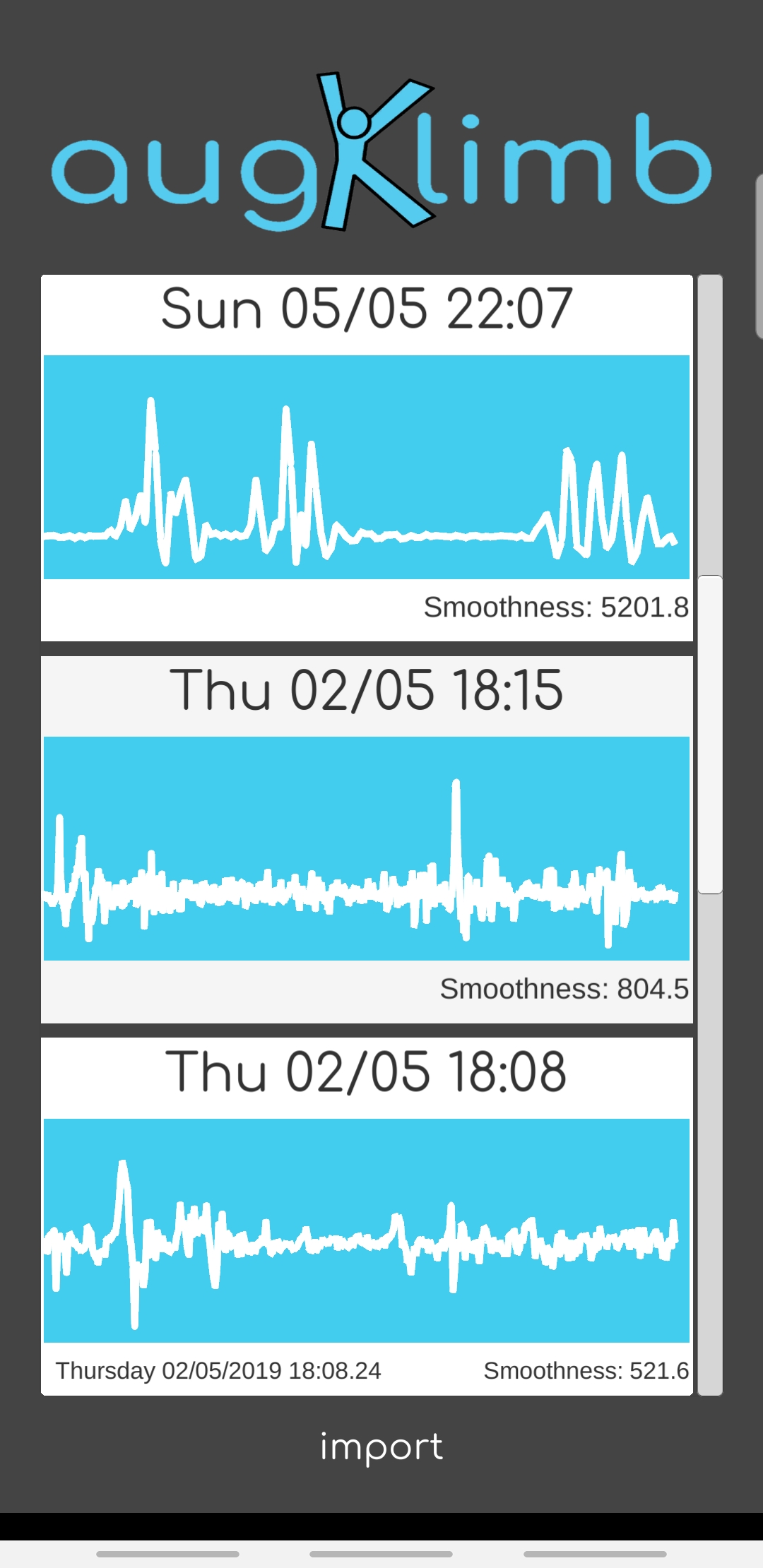}
\caption{The final version of augKlimb's single-climb-viewing page}
\label{fig:finalclimbview}
\end{figure}

To make a more coherent link between the overall smoothness-score and the graphs, I implemented the suggestion of calculating the per-second smoothness-score and including those on the graphs.
However, despite some users appreciating the added clarity of scores, some disliked the multitude of numbers everywhere, and felt that the graph was becoming very cluttered, with axis labels and lines and now scores, it was hard to see what was going on.
To aid with this visual, I vastly reduced the font-size of the smoothness scores, 
and instead portrayed the varying per-second score through changing the darkness of the background - see Figure~\ref{fig:finalclimbview}.

Users could now quickly scroll through the graph to the darkest section, and see which portion of their climb was causing the smoothness score to be high.

\subsection{Conflicting Requests From Users}
\label{conflict}
Due to the variety of use-cases the app can provide, a frequent side-effect of streamlining one user-flow was that other users would feed back that their preferred usage of the app had been hindered.

The biggest example of this was was which view to show the user upon completion of an accelerometer recording.
Some users who were using the graph to analyse their climbs, or wanted to quickly link a video recording, wanted that page to show up immediately after a recording had finished, rather than having to go back to view the lists of climbs and select that climb's page.
Others, who were just using the smoothness-score, only wanted to view that metric before being able to swiftly press the Record button again and continue climbing, and favoured the clean and simple UI that mode of operation provided, only occasionally going to look at the graphs to inspect weaknesses.

Differing proposals for improvements from both groups initially caused me to alter the app in each direction in turn, only when revisiting my changelog later did I notice that on a weekly basis I was switching back and forth between a very simplistic  UI and one that provided a lot of data, just dependant upon whose feedback I had been exposed to on those development and testing days.
To solve this, I invited three testers who fell across the above range, and prompted a discussion between them around both what they would expect to see, and what they wanted to see.
A compromise was formed, where the initial recording page showed nothing but the time elapsed and the smoothness score (see Figure~\ref{finalrecorderview}), with a shortcut link to that climb's specific page, which could contain the full list of statistics, graphs, editing abilities and video-linking facilities.

\begin{figure}[h]
\centering
\includegraphics[width=6cm]{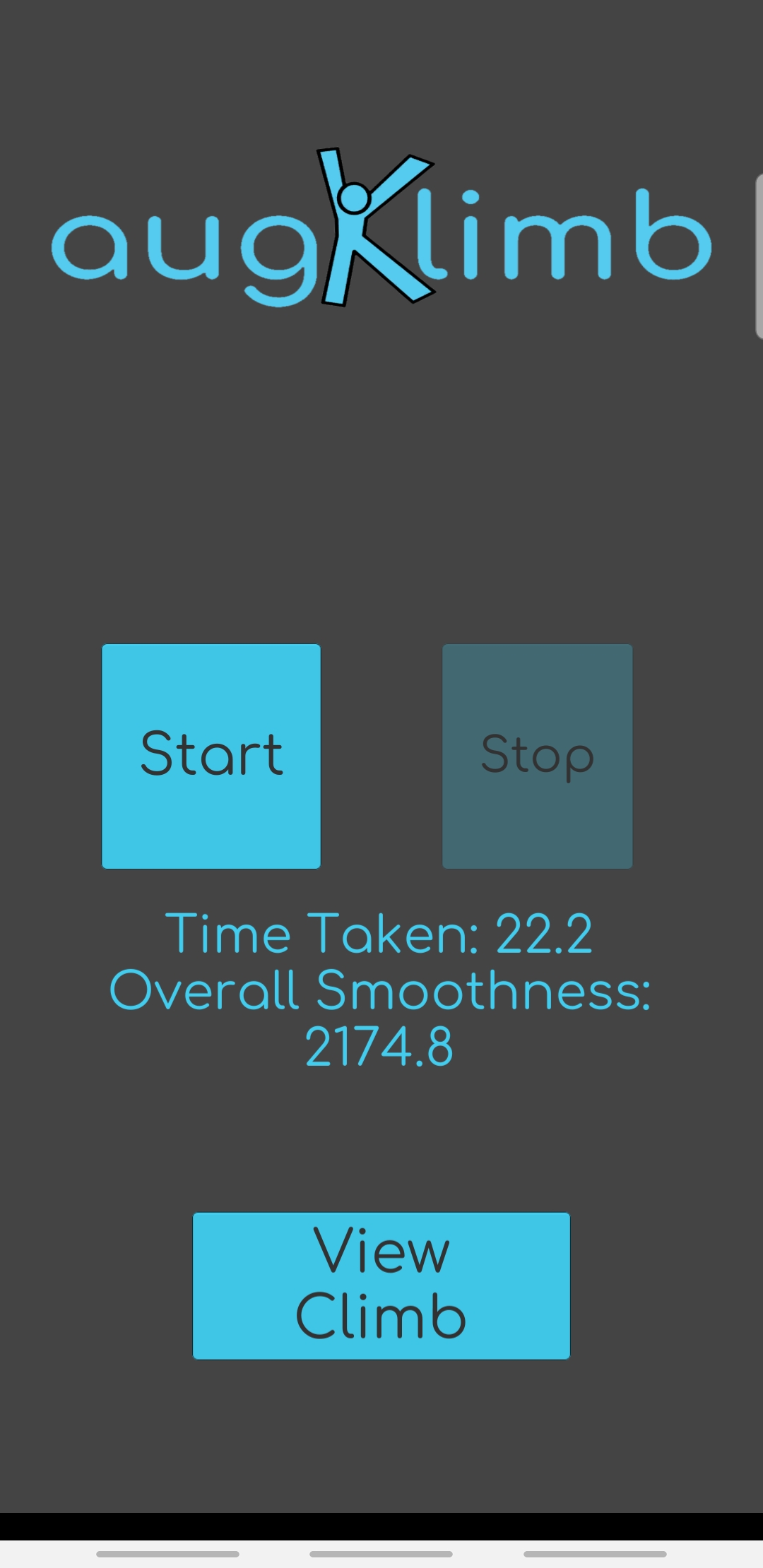}
\caption{The final version of augKlimb's post-recording view}
\label{fig:finalrecorderview}
\end{figure}

\chapter{Critical Evaluation}
\label{chap:evaluation}

Although the iterative design process can be seen as a never-ending spiral that continually refines a product, for the purpose of this project, I stopped iterating and deployed a ``final" version of the app.
This allowed me to spend the final two weeks assessing and evaluating, through a series of semi-structured interviews, both how well the app meets the original aims I set out to achieve as a product, and also to what extent my HCI-based hypotheses were accurate.

\section{Format of Final Evaluative Testing}
\subsection{Qualitative vs Quantitative}
Some of the previous research in the field has analysed the efficacy of their climbing products in a quantitative way, by statistically proving predictions of competition placements~\cite{climbaxstudy} or viewing an increase in climbing ability~\cite{climbbsn}. 
However, in the field of HCI, where usability and complex interactions between humans and technology are being examined, it is hard to extract this and reduce the data to numbers than can be statistically tested, leading to more qualitative methods being used~\cite{oro11911}.
Also, for the purposes of the project, recording progression for a quantitative analysis would require much more time than is available, and the hypotheses laid out at the start of the project require the deeper understanding that only qualitative analysis can provide.

\subsection{Thematic Analysis of Interview Data}
\subsubsection{Why this method}
Surveys were a useful source of information for both my initial requirements-gathering and for ongoing feedback points throughout the course of the development, but the static format inhibits elaboration (which leads to richer data), even with the most open-ended of questions~\cite{ozoksurvey}.

Therefore I opted to undertake a series of semi-structured interviews with a range of different users of the augKlimb app, and conduct a Thematic Analysis (TA) over the transcripts, a popular technique in recent HCI work~\cite{themanbrown}.

Semi-structured interviews are open discussions, guided by a set of rough topics or questions, but which allow for the exploration of new ideas by following any leads that come up throughout the conversation, producing rich data~\cite{oro11911}.

TA is a process of encoding qualitative information, and has been presented as the bridge between qualitative and quantitative methods~\cite{boyatzis1998transforming}
Thematic Analysis was developed in the field of psychology, and the most commonly cited methodology I found in HCI literature was the one laid out by Braun and Clarke~\cite{braunclarke06}.
They later wrote a chapter detailing how to apply thematic analysis to interview data, and so a later chapter~\cite{brauminterviewta} written by the same authors, detailing how to apply TA to interview data, was the guide I followed whilst performing the below analysis.

\clearpage
\subsection{Braun \& Clarke’s six phase approach to TA}
Here is the outline of the TA approach, detailed by Braun \& Clarke, that I performed:

\begin{quote}
\begin{enumerate}
    \item \textbf{Familiarisation with the data:} reading and re-reading the data.
    \item \textbf{Coding:} generating succinct labels that identify important features of the data relevant to answering the research question; after coding the entire dataset, collating codes and relevant data extracts.
    \item \textbf{Searching for themes:} examining the codes and collated data to identify significant broader patterns of meaning; collating data relevant to each candidate theme.
    \item \textbf{Reviewing themes:} checking the candidate themes against the dataset, to determine that they tell a convincing story that answers the research question. Themes may be refined, split, combined, or discarded.
    \item \textbf{Defining and naming themes:} developing a detailed analysis of each theme; choosing an informative name for each theme.
    \item \textbf{Writing up:} weaving together the analytic narrative and data extracts; contextualising the analysis in relation to existing literature.
\end{enumerate}

\hspace*{\fill}(Direct quote from~\cite{brauminterviewta})
\end{quote}

\section{Performing the Interviews}
\subsection{Question Selection}
The goals of my interviews were two-fold: to generally assess the usability and usefulness of the app, and also assess my hypotheses.
With this in mind, I developed a list of nine questions, which started by asking the climbers about their personal climbing (to both provide context and get the conversation flowing), and then lead on to how the app impacted their climbing, and how were their experiences of using it.

\begin{itemize}
    \item What do your climbing sessions usually include?
    \item Would you class your sessions as fun, training, or a mixture?
    \item How do you think using the app impacted your climbing session today?
    \item Do you see the app as a training tool or as gamification of (adding fun to) your climbs, or both?
    \item Do you think the app caused or helped improve your climbing?
    \item What was your favourite aspect?
    \item What was your least favourite aspect?
    \item Are you likely to continue using the app in the future?
    \item What feature(s) would you like to see extended or added?
\end{itemize}

These questions were only used as a guideline in the interviews, with more questions being asked to explore interesting points that were raised  throughout the conversation.
Although I aimed to obtain answers to all the questions, the wording and order of the questions varied, and often a participant would cover a question without prompting, during discussion leading from another question.

\subsection{Ethics}
Because this was a distinct user-study, with different aims and data-collection methods than my initial study, I applied for a second full ethics approval, the paperwork for which can be found in Appendix \ref{appx:ethics2}.
Privacy and anonymity concerns related to the recording of audio was the biggest challenge.
This was solved by transcribing the recorded audio files to text, before deleting the original recordings.
A secure list of names was kept to enable the particpants' right to withdraw, but any personally identifiable data was removed from the transcripts, and each participant was assigned a number, which is how I will refer to them throughout the below analysis.

\subsection{Participants}
I interviewed two males and four females, three of which were university students, and all were between the ages of 20 and 30.
It should be noted that this is not the most representative sample of the general climbing population, especially with regard to age, technology usage and climbing experience; many climbers are older and have been climbing for multiple decades.
Future research could explore how different age-groups or differently-experienced climbers interact with data-led augmentation of climbing.

\subsection{Transcription}
As stated above, after recording the interviews, I transcribed the audio into text, to enable the TA to be performed on the data.
Due to both privacy concerns and the inaccuracies of speech-recognition software, I manually transcribed the forty minutes of audio.
The full transcriptions from all six interviews can be found in Appendix~\ref{appx:transcriptions}.

Although this process was painstakingly laborious, and took over eight hours of typing to perform, it did also help to fulfil the first step of TA: to familiarise myself with the data.

\section{Thematic Analysis}
\subsection{Familiarisation}
Transcribing the audio helped begin my familiarisation with the data, but as this was a mostly passive process, trying to type the words as quickly as I could hear them, it did not provide me with the analytical immersion required for TA.
Therefore I repeatedly read through the transcripts, actively thinking about how the text applied to my research goals, and ``treating the data as \textit{data}"\cite{brauminterviewta} until I was fully engaged with the concepts highlighted through the discussion.

\subsection{Coding}
Leading on from the familiarisation, I began to systematically highlight and annotate the ideas and interesting points being raised by my interviewees. 
Where similarities arose, I ensured that the same annotation was applied consistently: these annotations were then \textit{codes}.

Multiple passes through the dataset was required as my analysis gradually became more developed, and more nuanced points were highlighted and re-discovered in previously annotated data.

\subsection{Theme Identification}
Patterns gradually arose in the codes I was noting down, also known as ``themes" in the TA terminology. 
I kept these themes in mind as I read through the data set again, colour-coding codes that fell within the various meanings.
After reviewing the themes, and the various codes that they collated, against the transcripts again, I defined and named them.

The themes that arose, and a brief selection of the codes that characterised them, can be found in Table~\ref{tab:themes}.

\begin{table}[h]
\begin{tabular}{|l|l|}
\hline
Route-difficulty     & 
warming-up with easy climbs, technique focus on easy climbs,
\\& repeating the same climb, attempts to climb at limits, 
\\& app more useful on easy climbs
\\ \hline

Seriousness of a climbing session   &  
fun, training, gamification, social interaction, competition with self,
\\& competition with others
\\ \hline

Complexity of analytics provided    & 
quick performance feedback, detailed analytic feedback,
\\& simple score, graph spikes, video to graph visualisation,
\\& request for labelling technique as "good or bad"
\\ \hline

Ease of use                   & 
easy to use, simple, request for more instructions, 
\\& not wanting to look at the screen, too many button presses,
\\& difficulty in connecting video, gui output is good,
\\& linking repeated attempts of climbs
\\ \hline

Mobile phone as a form-factor       & 
suggestion for wristband, instant display useful, lack of pockets
\\ \hline

\end{tabular}
\label{tab:themes}
\end{table}

These themes align quite closely with both my hypotheses and with the questions asked in the interviews.
There are two reasons for this: the interviews were conducted with guideline questions, so the discussions prompted from these guidelines will naturally follow similar topics, and also I had my research aims in the back of my mind whilst performing both the coding and theme-collation steps, as recommended by the guide I was following~(\cite{brauminterviewta}).

\subsection{Discussion}
I will now discuss what I learnt from this analysis, first in relation to each of the four hypothesis points, and then on a theme-by-theme basis.

\subsubsection{Relation to Hypotheses}
Here are the four hypotheses laid out in Chapter~\ref{chap:context}:
\begin{enumerate}
    \item Augmenting a climbing session with a live-feed of data analytics will positively impact climbing technique.
    \item A lightweight and simple-to-use product will be popular among intermediate climbers who are serious enough to want to improve, but not so serious they want to pay for coaching.
    \item Seeing a ``score" that rates climbing technique will enable gamification and fun, both for individuals and within groups.
    \item Providing more data to climbers will enable more focused progression tracking and goal-oriented training.
\end{enumerate}

All of these hypotheses were at least partially-confirmed during the testing:
\begin{enumerate}
    \item Although the first hypothesis may have lent itself to a more quantitative analysis, I had to rely on the climbers' self- perception on whether their climbing improved. Although P6 did not feel as though the app had impacted their ability, all the other interviewers said it had some form of impact: either in the short-term (for example P5 saying that the ``pressure" from knowing they were being recorded made them ``think a lot more" on smooth technique) or the long term (P1 saying they were ``climbing better" after using the app).
    \item With all of the interviewees being within my target demographic, all saying they enjoyed using the app, and four out of six of the interviewees saying that they are definitely going to continue using the app in the future, it can be concluded that the app is popular among the type of climbers I was aiming to develop it for.
    \item The smoothness score definitely enabled fun, with P3 saying that they ``loved the gamification, wanting to get the scores as high or as low as possible, that is quite good fun", yet some interviewees who saw the app as more of an analytics tool ``hadn’t really considered treating it like a game to try and get a better score"(P1).
    \item P2 in particular enjoyed the more analytic progression-teacking available, using the app mainly as a ``figure for performance" and to ``compare the two times I've managed to do those climbs". However, there were some limitations with the app's ease of use in this area, with P3 saying that although it was ``interesting to see how they compare" when tracking progression, it was ``easy to mix it up" when trying to scroll back and view a previous attempt at a climb.
\end{enumerate}

\subsubsection{Route-difficulty}
A variety of comments were made in the interviews about how differently graded routes were deliberately climbed at different times throughout a session. 
The average session seemed to consist mostly of ``warming up at the beginning with easier climbs, and then working up towards harder climbs"(P1).

Across all participants, the app was used more often to focus on smoothness during easier climbing, partly because it was ``a very easy way to use the app"(P4).

Interestingly, for some interviewees, choosing to use the app seemed to impact the choice of route-difficulty, whilst conversely for others, the choice of climb impacted how the app was being used:
When P3 decided to spend time using the app, they would select an easy route and ``keep doing the same climb" until they got the score as low as possible.

Alternatively, P4 stated that ``when I'm warming up, I'll be looking at my technique, trying to do things slowly and smoothly, like that's when I'd really be looking at the app" to quickly determine smoothness score.
Then, they ``try to remember those smooth movements when I move onto the harder climbs", but if they fall off or get stuck, they often ``wanted a full analysis with the video as well" to help them determine the weak points.

Whenever these two different use-cases were mentioned, they were linked to the perceived difficulty of the route being attempted.

\subsubsection{Seriousness of a climbing session}

The interviewees had a range of views about whether their climbing sessions were for fun or training. 
Both P3 and P5 said their sessions were always just fun-orientated, 
P1 and P4 had similar views in that they climbed ``mostly for fun but obviously ... I want to keep improving"(P1) and that ``the more you train the more fun it gets"(P3), and P6 deliberately alternated, doing ``two training sessions and one fun session" per week, and P2 focusing almost entirely on the ``training side".

This had a strong correlation with how they perceived the app, as either a training tool or as a gamifying aid to fun. 
Gamification has already been discussed above in the analysis of the third hypothesis, so I won't repeat it here.
The ability of the app to be used in the same way by different people (in that they click, climb, and interpret the data identically), but then for them to emotionally respond in such a different manner, as either a competitive fun element or as a drive to improve their performance, depending upon what their goals were for the session, was an interesting result.

\subsubsection{Complexity of Analytics}

The two different analytical uses of the app, covered both in Section~\ref{conflict} and briefly in the Route-Difficulty theme, were interesting.
Despite a wide range of different ways the app could be used, all interviewees described either: (1) quickly comparing just the smoothness score of multiple ascents, or (2) going into a detailed analysis with the graphs and video to determine why they struggled with a harder route.
This suggests that users want to interpret very complex analytics, or quickly view very simple analytics, but not anything in the middle.

Also grouped within this theme were the codes that related to complaints and improvement-suggestions relating to the analytics sometimes being undecipherable, with P5 saying they didn't enjoy the smoothness score because they ``didn't really have an idea of where the scale went from and to on the rating", and P3 suggesting that using a 1`simpler 5-star rating" would be easier to use, as being less experienced, they ``don't really know what some of the stats mean".

This was good feedback, and I intend to implement the suggestions on future versions of the app, but as I will elaborate on in Section~\ref{aimsconf}, the test version of the app didn't contain any guidelines, to allow my to examine how climbers naturally use those datapoints when given no prompting.

\subsubsection{Ease of use}

The app was generally stated as ``pretty easy to use"(P2) by all interviewees, most ``liked the interface``(P6), and P5 even stated the ease-of-use as their favourite aspect. However, some aspects and use-scenarios were found to be more difficult or not slick enough.

As detailed above, a general guide on how to use the app would have been appreciated by most participants.

P2 disliked how, because they jumped off the wall at the end of a climb, they had to ``delete the spike from the jumping" before an accurate smoothness score was shown, as ``it felt like there were too many buttons that I had to press every time", and suggested an automatic-detection of such a fall to speed up and ease how they used the app

The linking of devices, to transfer video or climb-files was the most commonly reported issue, with it being ``quite complicated"(P1) to do so, a problem I explored in depth in section~\ref{network}, but due to the limitations of Unity, was unable to find a more satisfactory solution to.

\subsubsection{Mobile phone as a form-factor} 

Although they appreciated the use of a mobile phone for displaying the output in an interactive way, most of the interviewees highlighted the device's shortcomings as a data-collection tool.
Particularly P6, who stated the phone as their least favourite part of the app, because ``I don't have any pockets on my clothing, so I find it quite hard to put it somewhere".
They then went on to say ``what I'd like to see in the future is linking it to a wristband or a smartwatch or something", an idea echoed by P4, who said ``it would be great to have external sensors", and P2, who didn't like how a phone ``wobbled`` in their pocket, and would prefer it ``if it was on an armband".

P4 did actually use the phone with it attached to their body with an arm-strap, but said that the movements were sometimes ``too jolty", and next time wanted to either ``wear pockets ... to see what results it gives when I've got it attached to my torso", or ``putting it on difficult limbs, to see if you're working harder on each hand for instance", interesting points I had not considered.

\chapter{Conclusion}
\label{chap:conclusion}

\section{Main Achievements}
The main achievement of this project was the development of the \verb|augKlimb| app.

My iterative user-centric process involved over $60$ hours of in-field observations of (and discussions with) climbers interacting with various prototypes, culminating in a market-ready app that is published and deployed online.

For a final evaluation, I interviewed six climbers who had been using the app, and then conducted a thematic analysis of the interview transcripts, gaining a deep understanding of the app's impact on the climbers using it, along with its various use-cases, benefits and limitations.

\section{Project Status}
\subsection{App developed}
Iterative design never truly ends, and the series of interviews I conducted at the end of this project highlighted many potential directions I could take the app in the future, which I intend to continue doing even after the completion of this project - I go into more detail below.
However, in order to conduct that final usability study, I drew a line and deployed a fully-working version of the app, which is on the Google Play store, and is currently being used on a regular basis by around 15 local climbers, something I am very proud of.

\subsection{Features}
The app can can record the acceleration of a climber, displaying it as a graph, and annotate second-by-second smoothness ratings.
A video can be optionally linked to the graph, providing context to the peaks and troughs, and enabling frame-by-frame playback of the climb.

Although these are good features, and allow both deep and shallow levels of technique feedback, they are not as technically advanced as I was hoping to achieve when I first started this project.
By restricting myself to a mobile phone as the device being used (a deliberate choice prompted by both my initial survey and my want for the final product to be as accessible as possible), I was limited to only 2D video and 1D accelerometer data as possible inputs.
In my rush to get a coded prototype out to testing, I didn't spend a lot of time experimenting with video-based analysis techniques, but built an accelerometer-based app, planning to revisit the possibility of CV at a later date.
Inevitably, as my I iterated through the development, effectively adding extra visualisations and analytic each time, the core essence of the app remained centred on the accelerometer as the primary method of input.

Perhaps if I were to do this project again, I would spend more time at the beginning developing a CV-based analysis feature, which just feels more interesting and innovative than graphs and statistics based on acceleration data. 
However, I still believe that in the context of a user-centred design, and given the time constraints of the project, I made the right choices given what I knew at the time.

\subsection{Future Plans}
\subsubsection{Short-Term}
The interviews and subsequent analysis acted as part of the Iterative Design cycle, so my immediate inclination is to act upon the easier-to-implement suggestions, and improve the app by adding an introductory guide, the ability to link data from repeated attempts at the same climb, and more connectivity over social media to enhance remote competition.

\subsubsection{Mid-Term}
The unexplored option of CV-analysis, potentially aided by cloud-computing, would be an interesting area for future development.
Especially as the last major iterative loop was to add video-playback to the app, recording and importing videos are a part of the current user-pattern, and so gradually starting to use these to produce more statistics would be a good idea.

One of the major issues with the current app is the difficulty in transferring and importing the video or climbing files between devices.
This was mainly caused by the usage of Unity as the app-development tool.
Although Unity was great for the quick iteration and testing of a simple app, now I have a more refined idea of what the app's features and requirements are, I could potentially port it over to another platform that doesn't offer as much speed or iterative support, but instead offers better device-connectivity options.
This would also potentially open up the ability for the app to interface with wearables or smartwatches, a much-requested feature that was not possible with Unity.

\subsubsection{Long-Term}
When it comes to a more general direction in which to take the app, two potential suggestions came out of the interviews:
\begin{itemize}
    \item Increase the analytical component of the app - add goals, technique drills and plans, long-term progression tracking, and other training-oriented features.
    \item Accentuate the capacity for gamification - add more social features, maybe even moving towards turning the app into a literal game, with points scored per climb and online connectivity with friends.
\end{itemize}

\section{Comparison to Original Aims}
My general aim was to iteratively build a working and useful product, and then analyse how the data given to climbers was used, with a set of hypotheses.

I managed to successfully meet both of these aims, although at times they had slight conflicts. \label{aimsconf}
If developing the best app possible had been my only goal, then I would have included guides on how to interpret the data and examples of how the app can be used in different ways.
However, this set of instructions would have severely limited my ability to analyse how the climbers themselves interpreted the data, and so the version of the app I used for the final study did not include them, which caused issues as were highlighted through the ``Ease-Of-Use" and ``Complexity of analytics" themes in the TA.

From my original proposal, my more detailed aims were as follows:
\begin{itemize}
    \item Must Have:
    \begin{itemize}
        \item  At least 2 wizard-of-oz prototypes to test user interaction and preferences.
        \item A final product that implements some of the features at a low-fi level.
        \item  Some form of testing and analysis of both prototypes and the final product.
        \item Full ethics and health-and-safety approval.
    \end{itemize}
    
    \item Should Have:
    \begin{itemize}
        \item An analysis of the current needs and wants of intermediate-level climbers.
        \item A final product that fully works and is very useful for intended purpose.
        \item Plenty of user testing to give both qualitative and quantitative results for both the prototypes and final design.
    \end{itemize}
    
    \item Could Have: 
    \begin{itemize}
        \item Strong usage of User-centred-design techniques.
        \item Usage of a very novel technology/technologies as part of the final product.
        \item A final product that is ready for market.
        \item An excellent writeup analysing choices and mistakes made along the development process.
    \end{itemize}
\end{itemize}

The Must-Haves were very conservative, so I easily met those aims.

I partially met all the Should-Haves: whether the final app is \textit{very} useful is a subjective matter I am not entirely sure about, it is certainly useful in some ways, but also has its limitations.
I conducted a lot of qualitative user-testing, but the closest I got to using quantitative data was for the development of my smoothness score.

Moving onto the Could-Have aims: I definitely followed a very user-centred-design-centred methodology; my main source of dissatisfaction with the project is the lack of a very novel technology in the final product - as explained above, I prioritised user-centrism over novelty of features; my final product is not only ready for market, it is in the market and being actively used, yet can definitely be improved-upon, especially after my deeper analysis of its limitations.

As for the final Could-Have, I have attempted to analyse the choices and mistakes made throughout my project, but I leave it to the reader to decide whether this writeup is excellent or not.

\clearpage

\bibliographystyle{plain}
\bibliography{refs}

\begin{thebibliography}{10}

\bibitem{oro11911}
Anne Adams, Peter Lunt, and Paul Cairns.
\newblock A qualititative approach to hci research.
\newblock In Paul Cairns and Anna Cox, editors, {\em Research Methods for
  Human-Computer Interaction}, pages 138--157. Cambridge University Press,
  Cambridge, UK, 2008.

\bibitem{ozoksurvey}
A~Ant~Ozok.
\newblock Survey design and implementation in hci.
\newblock pages 151--1169, 03 2009.

\bibitem{bacafeedback}
A.~{Baca} and P.~{Kornfeind}.
\newblock Rapid feedback systems for elite sports training.
\newblock {\em IEEE Pervasive Computing}, 5(4):70--76, Oct 2006.

\bibitem{socialclimb}
Natalie Berry.
\newblock Social climbers - the evolving indoor climbing industry.
\newblock {\em ukclimbing.com}, July 2018.

\bibitem{boyatzis1998transforming}
R.E. Boyatzis.
\newblock {\em Transforming Qualitative Information: Thematic Analysis and Code
  Development}.
\newblock Transforming Qualitative Information: Thematic Analysis and Code
  Development. SAGE Publications, 1998.

\bibitem{pedometer}
Glenn Boyce, Gayan Padmasekara, and Martin Blum.
\newblock Accuracy of mobile phone pedometer technology.
\newblock {\em Journal of Mobile Technology in Medicine}, 1:16--22, 06 2012.

\bibitem{braunclarke06}
Virginia Braun and Victoria Clarke.
\newblock Using thematic analysis in psychology.
\newblock {\em Qualitative Research in Psychology}, 3(2):77--101, 2006.

\bibitem{themanbrown}
Nela Brown and Tony Stockman.
\newblock Examining the use of thematic analysis as a tool for informing design
  of new family communication technologies.
\newblock In {\em Proceedings of the 27th International BCS Human Computer
  Interaction Conference}, BCS-HCI '13, pages 21:1--21:6, Swinton, UK, UK,
  2013. British Computer Society.

\bibitem{callawayvideoacccomp}
Andrew~J Callaway, Jon~E Cobb, and Ian Jones.
\newblock A comparison of video and accelerometer based approaches applied to
  performance monitoring in swimming.
\newblock {\em International Journal of Sports Science \& Coaching},
  4(1):139--153, 2009.

\bibitem{chalkprint}
chalkprint.
\newblock Comprehend your climbing.
\newblock \url{https://chalkprint.com/}.

\bibitem{automaticrowingcoach}
Simon Fothergill, Robert Harle, and Sean Holden.
\newblock Modeling the model athlete: Automatic coaching of rowing technique.
\newblock In {\em Proceedings of the 2008 Joint IAPR International Workshop on
  Structural, Syntactic, and Statistical Pattern Recognition}, SSPR \& SPR '08,
  pages 372--381, Berlin, Heidelberg, 2008. Springer-Verlag.

\bibitem{sportperformance86}
Ian~M. Franks and David Goodman.
\newblock A systematic approach to analysing sports performance.
\newblock {\em Journal of Sports Sciences}, 4(1):49--59, 1986.
\newblock PMID: 3735484.

\bibitem{edgeinteractive}
Andre Kennedy Christoph~Zobl Franziska~Heuck, Aileen~Kassing.
\newblock Edge | an interactive training wall for climbers.
\newblock
  \url{https://designawards.core77.com/interaction/51284/edge-an-interactive-training-wall-for-climbers}.

\bibitem{groomcoachperceptions}
Ryan Groom.
\newblock Coaches perceptions of the use of video analysis.
\newblock {\em Insight}, 7, 08 2004.

\bibitem{groomvideo}
Ryan Groom and Christopher Cushion.
\newblock Using of video based coaching with players: A case study.
\newblock {\em International Journal of Performance Analysis in Sport},
  5:40--46, 12 2005.

\bibitem{ISO9241-210}
{Ergonomics of human-system interaction -- Part 210: Human-centred design for
  interactive systems}.
\newblock Standard, International Organization for Standardization, Geneva, CH,
  March 2010.

\bibitem{projectedclimbwall}
Raine Kajastila and Perttu H\"{a}m\"{a}l\"{a}inen.
\newblock Augmented climbing: Interacting with projected graphics on a climbing
  wall.
\newblock In {\em Proceedings of the Extended Abstracts of the 32Nd Annual ACM
  Conference on Human Factors in Computing Systems}, CHI EA '14, pages
  1279--1284, New York, NY, USA, 2014. ACM.

\bibitem{climbaxstudy}
Cassim Ladha, Nils Hammerla, Patrick Olivier, and Thomas Ploetz.
\newblock Climbax: Skill assessment for climbing enthusiasts.
\newblock 09 2013.

\bibitem{lieberreview}
Dario~G. Liebermann, Larry Katz, Mike Hughes, Roger M~Bartlett, Jim McClements,
  and Ian Franks.
\newblock Advances in the application of information technology to sport
  performance.
\newblock {\em Journal of sports sciences}, 20:755--69, 11 2002.

\bibitem{cvinsport}
Thomas~B. Moeslund, Graham Thomas, and Adrian Hilton.
\newblock {\em Computer Vision in Sports}.
\newblock Springer Publishing Company, Incorporated, 2015.

\bibitem{niederervis}
C.~{Niederer}, A.~{Rind}, and W.~{Aigner}.
\newblock Multi-device visualisation design for climbing self-assessment.
\newblock In {\em 2016 20th International Conference Information Visualisation
  (IV)}, pages 171--176, July 2016.

\bibitem{kinovea}
{Nor Muaza} {Nor Adnan}, {Mohd Nor Azmi} {Ab Patar}, Hokyoo Lee, {Shin Ichiroh}
  Yamamoto, Lee Jong-Young, and Jamaluddin Mahmud.
\newblock Biomechanical analysis using kinovea for sports application.
\newblock {\em IOP Conference Series: Materials Science and Engineering},
  342(1), 4 2018.

\bibitem{odonovideo}
Peter O’Donoghue.
\newblock The use of feedback videos in sport.
\newblock {\em International Journal of Performance Analysis in Sport},
  6(2):1--14, 2006.

\bibitem{oghiswim}
Y.~{Ohgi}.
\newblock Microcomputer-based acceleration sensor device for sports
  biomechanics -stroke evaluation by using swimmer's wrist acceleration.
\newblock In {\em SENSORS, 2002 IEEE}, volume~1, pages 699--704 vol.1, June
  2002.

\bibitem{pansiottenniscv}
J.~{Pansiot}, A.~{Elsaify}, B.~{Lo}, and {Guang-Zhong Yang}.
\newblock Racket: Real-time autonomous computation of kinematic elements in
  tennis.
\newblock In {\em 2009 IEEE 12th International Conference on Computer Vision
  Workshops, ICCV Workshops}, pages 773--779, Sep. 2009.

\bibitem{climbbsn}
J.~{Pansiot}, R.~C. {King}, D.~G. {McIlwraith}, B.~P.~L. {Lo}, and {Guang-Zhong
  Yang}.
\newblock Climbsn: Climber performance monitoring with bsn.
\newblock In {\em 2008 5th International Summer School and Symposium on Medical
  Devices and Biosensors}, pages 33--36, June 2008.

\bibitem{strangebeta}
C.~{Phillips}, L.~{Becker}, and E.~{Bradley}.
\newblock strange beta: An assistance system for indoor rock climbing route
  setting.
\newblock {\em Chaos}, 22(1):013130, Mar 2012.

\bibitem{schmidt2005motor}
R.A. Schmidt and T.D. Lee.
\newblock {\em Motor Control and Learning: A Behavioral Emphasis}.
\newblock Human Kinetics, 2005.

\bibitem{schmidt75aschema}
Richard~A. Schmidt, Ronald~G. Marteniuk, and Karl~M. Newell.
\newblock A schema theory of discrete motor skill learning.
\newblock {\em Psychological Review}, pages 225--260, 1975.

\bibitem{centreofmass}
F.~Sibella, I.~Frosio, F.~Schena, and N.A. Borghese.
\newblock 3d analysis of the body center of mass in rock climbing.
\newblock {\em Human Movement Science}, 26(6):841 -- 852, 2007.

\bibitem{lattice}
Lattice Training.
\newblock About lattice training.
\newblock \url{https://latticetraining.com/about-us/}.

\bibitem{verticallife}
VerticalLife.
\newblock Vertical-life climbing.
\newblock \url{https://www.vertical-life.info/}.

\bibitem{brauminterviewta}
Victoria~Clarke Virginia~Braun and Nicola Rance.
\newblock How to use thematic analysis with interview data.
\newblock In {\em The Counselling and Psychotherapy Research Handbook}, pages
  183--197. SAGE Publications Ltd, 55 City Road, London, 2015.

\bibitem{climbing-sub-worlds}
Brandon Wayne~Rapelje.
\newblock Rock climbing sub-worlds: a segmentation study.
\newblock 08 2004.

\bibitem{whipper}
Whipper.
\newblock Whipper: World’s 1st climbing performance tracker.
\newblock
  \url{https://www.indiegogo.com/projects/whipper-world-s-1st-climbing-performance-tracker#/}.

\end{thebibliography}

\appendix

\chapter{Interview Transcriptions}
\label{appx:transcriptions}
\section{Participant One}
LS:  Interview with Participant One. Hello!
\\ P1:  Hi
\\ LS:  So, what do your climbing sessions normally include?
\\ P1:  Um, my climbing sessions usually involve about half an hour of warming up, stretching and doing some easier climbs, then maybe another couple of hours of bouldering, like trying to do new routes and going over some of the ones that I didn't do quite right last time. And then obviously some stretching at the end - very important.
\\ LS:  Would you class your sessions as fun, training, or a mixture of the two?
\\ P1:  I would say it's mostly for fun but obviously I want to get better, I want to keep improving
\\ LS:  How do you think using the app impacted your training session today?
\\ P1:  Using the app made me do the same climb more than once, and improve on the climb each time that I did it, trying to get rid of some of the unnecessary dynamic moves
\\ LS:  Which feature helped with that in particular?
\\ P1:  The accelerometer, and also the video, like watching and seeing which moves were maybe too big so I could work on them.
\\ LS:  Ok, cool, which would you say was your favourite aspect or feature of the app?
\\ P1:  I would say that the video, being able to see and watch what impact it had on your...
\\ LS:  Oh, so the interplay between the video and the accelerometer data?
\\ P1:  Yeah, exactly!
\\ LS:  So how does the video feedback on the app compared to just having video playback by itself, without the app? More or less useful?
\\ P1:  Just video, I think, would be less useful, because, watching yourself back. If you've just done the climb and you don't know which bits were more or less dynamic, you could watch it back and think that you'd done really well, whereas really you could have been doing a lot better.
\\ LS:  So you mean like, scrolling through the graph to see the colour coding and the peaks, and then where that links up with the video-d moves?
\\ P1:  Yeah I liked that a lot.
\\ LS:  So what would you say was your least favourite aspect of the app?
\\ P1:  I didn't really know how to link or send the video stuff between phones, it seemed quite complicated.
\\ LS:  Yeah it can be a bit of faff to get the Bluetooth working...
\\ P1:  (laughs)
\\ LS:  Are you likely to continue using the app in the future?
\\ P1:  Yeah, if you could help show me how to do that transferring of files stuff again
\\ LS:  (laugh) Yeah I'll add some more detailed guides on the website and the app I guess.
\\ P1:  are there any reasons you wouldn't bother using the app in the future?
\\ P1:  Um, if I was just doing fun climbing I probably would use the app again, but if I wanted to focus on getting better then I would use it yeah
\\ LS:  So do you think that the app helped improve your climbing?
\\ P1:  Yeah, I think so, having to do the same climb more than once to improve the score meant that you were climbing better, with more focus on technique.
\\ LS:  Would you say that you see the app as a training tool or more as a game, seeing a score of the climbs?
\\ P1:  I would say its more of a training tool, but then I hadn't really considered treating it like a game to try and get a better score 
\\ LS:  OK, I guess I'm asking like how much does the app add fun to your climbs, as in is it an analytics tool to you, or is it a fun-making exercise to use the app?
\\ P1:  I'd go with the first one, that it's an analytics tool to see my climbs
\\ LS:  Ok, so which feature would you like to see added or extended in the future?
\\ P1:  I don't really know.
\\ LS:  If you had to come up with an improvement, what would you say?
\\ P1:  Maybe making the video more easy to connect in?
\\ LS:  Yeah fair point haha. Any other comments you'd like to make?
\\ P1:  I don't think so. Sick job!
\\ LS:  (laugh) cheers

\section{Participant Two}
LS:  Alright, interview with participant number two. First question, what do you climbing sessions normally involve?
\\ P2:  It involves warming up at the beginning with easier climbs, and then working up towards harder climbs, and then after that probably a little bit of a workout, like pullups, then a cool down yeah
\\ LS:  so would you class that normal session as fun or more as a training session?
\\ P2:  Something in between. but more on the training side
\\ LS:  How do you think using the app has impacted your session today?
\\ P2:  I feel like it can provide good feedback on performance...
\\ LS:  ok yeah, go on
\\ P2:  and like basically a figure to compare daily  sessions against each other, perhaps
\\ LS:  so would it be interesting to come back next time and compare to see how the numbers are for doing the same climbs?
\\ P2:  yeah exactly
\\ LS:  So do you see those number as a training thing then?
\\ P2:  yeah, for me, yeah. So, for instance, if I try something and I am not able to do it today, then next time I come and I do it, then the session after that, I can come and do it, and compare the two times I've managed to do those climbs. 
\\ LS:  What in particular would you be comparing in that case?
\\ P2:  The smoothness score mostly
\\ LS:  So what would you say your favourite feature of the app was?
\\ P2:  Yeah that smoothness score was what I used the most
\\ LS:  How did you find the video feature?
\\ P2:  It was really good actually, yeah. It was to combine the two, seeing the movements with the video and so you can actually see the accelerations of all of them and see when they happened
\\ LS:  The interaction between the two?
\\ P2:  yeah
\\ LS:  Do you see the app overall as a fun, almost gamification of the climbing, or a training tool to set goals or targets?
\\ P2:  I wouldn't use it to set targets, but I would use it more as a figure for performance. Not a target I want to achieve for smoothness, but more as a figure, you know, as a figure to compare my climbing daily, what my smoothness is.
\\ LS:  What was your least favourite aspect of the app?
\\ P2:  The fact that at the end of the climbing I had to delete the spike from the jumping, that cropping before seeing the score.
\\ LS:  Yeah, OK, would that be your main suggestion for improvement?
\\ P2:  Yeah I think we can do some improvement, but I think when you link the video with the acceleration data, then you can see when stuff is happening, which is much easier, but it you just look at the graph data by itself then it can be a bit difficult to see what is happening. I would also say that if someone is using it regularly, then maybe it could have an armband, or like an footband or something, so the location be stationary, so you can compare sessions with each other. Whereas if it's in your pocket then the location it can just wobble around and can be different each time.
\\ LS:  Yeah a wristband could be a good way of using it, like one of those ones runners have?
\\ P2:  Yeah just for consistency of where it is located on your body.
\\ LS:  Sure yeah, good point. Are you likely to continue to use the app in the future after this study, for your own climbing.
\\ P2:  I would say yes, but I would also say that if I wanted to use it then it should be easier to use. Like at the moment it felt like there were too many buttons that I had to press, and every time, take it out of my pocket and do it. Maybe like an easier interface or something. I mean, it is pretty easy to use, with just the two buttons, but...
\\ LS:  Is it that cutting off of the jump down that you don't like?
\\ P2:  Yes.
\\ LS:  So like if you could just press stop and straight away see the smoothness?
\\ P2:  Yeah, maybe if it was on an armband , so I didn't have to look at the screen, I could just press the buttons, and feel vibrations when it records start and stop.
\\ LS:  Nice, yeah. Then if you don't want to be looking at a screen, what output would you want? 
\\ P2:  Yeah maybe, I like the smoothness, but like I said, maybe not right then. If I'm comparing it over two days then afterwards I can just bring up those two specific routes and have a look a them on the phone.
\\ LS:  Yeah, fair.
\\ P2:  So yes basically if I am doing a route for an instance, it has another button I can press and it goes into favourites, and at the end of the session I can name it maybe. And then next time I come, I can go like, ok this is the one, I am trying to do it again, and can compare them with each other.
\\ LS:  Cool, yeah, nice idea! Any further comments?
\\ P2:  Nah, good work man!
\\ LS:  Thanks!

\section{Participant Three}
LS:  This is the interview with participant three.
\\ P3:  Hi!
\\ LS:  What do your climbing sessions usually include?
\\ P3:  A bit of a warm-up, then I do some slightly more difficult climbs that I still know I can do, and then I'll try some I'm not sure I can do or not, and then I'll end with a bit of an overhanging sessions, and then that's done.
\\ LS:  OK, would you class your normal sessions as just fun, or more training-orientated?
\\ P3:  Fun
\\ LS:  How do you think using the app has impacted your climbing session?
\\ P3:  It's made me think a bit more about when I'm climbing, what I'm doing. It also makes me slow down a little bit, and actually take the time to take my phone out of my pocket and actually just consider what I'm about to do.
\\ LS:  Ok, so you've used the app on multiple sessions now, have you ever compared scores between sessions?
\\ P3:  Yeah I have
\\ LS:  And how has using the app on successive sessions impacted how you use it?
\\ P3:  Being able to name the sessions, so you know which is which, was a useful addition that made that easier to do, but I'm not really going climbing for training, but it was interesting to see how they compare. It wasn't always easy to do though.
\\ LS:  maybe an easier way of linking a climb with a previous one would be good then?
\\ P3:  Yeah, or even maybe assigning a photo, because with just writing a description, it is still easy to mix it up
\\ LS:  What about the video?
\\ P3:  Yeah that's a lovely feature, but I'd need someone to record it, and I usually climb alone
\\ LS:  What was your favourite aspect of using the app?
\\ P3:  The stats. I loved the gamification, wanting to get the scores as high or as low as possible, that is quite good fun, because its like, ok how can I be better
\\ LS:  ok yeah, so is that against yourself or others?
\\ P3:  Just against myself by repeating a climb or between climbs yeah
\\ LS:  What was your least favourite aspect of the app?
\\ P3:  Having to stop and start each time, and having to... I guess I'm just really bad at turning the screen off by accident, or pressing stop when it's in my pocket
\\ LS:  So if you had to suggest an improvement to the app, would that be it?
P3:  I guess the easiest improvement would be to add a key, or give  a simpler star rating or something, because saying is a high number good or bad, or what is expected for a smooth or dynamic climb's score. Being a bit of a noob climber, I don;t really know what some of the stats mean.
\\ LS:  Yeah, I suppose the idea was for the climber can decide how to use it, and what the numbers mean
\\ P3:  I guess just giving more of a definition, like this is what this score means
\\ LS:  I had a question about whether you see the app as more of a gamification or as a training tool, but I think you've already covered a bit of that...
\\ P3:  I like the idea of, this is what I achieved last session, lets see what I can do next session
\\ LS:  Do you think the app has caused to helped improve your climbing at all?
\\ P3:  Yeah, it's definitely made me think more about it, and how I can try to lower the numbers.
\\ LS:  What specifically has it made you think more about?
\\ P3:  I guess, technique, trying to think about what I can do to increase my smoothness, and be less dynamic, and be a bit more controlled.
\\ LS:  So have you done drills effectively, using the score?
\\ P3:  I just keep doing the same climb, there were a few times when the centre was quite empty and I could just do that
\\ LS:  Nice. Any features you'd like to see extended, or added?
\\ P3:  I suppose a bit more analysis... actually I dunno... It could be kind of cool if it said like "I think you did this here", but even then.. bouldering is too short for that. Actually yeah, more gamification stuff could be cool, a leaderboard, ways so you could be like "ha I'm better than you"
\\ LS:  so to bring in more actual like social competition between users
\\ P3:  yeah
\\ LS:  Any further points you'd like to make?
\\ P3:  Nope. I really like the idea, its a really good app!
L3: Cool, thank you very much

\section{Participant Four}
LS: Interview with participant number four. Hello!
\\ P4: hello!
\\ LS: What do your climbing sessions usually involve?
\\ P4: Um, a warm-up, very important, a few kind of easier climbs that I know I can do fairly well, and then start to build up from there. I Usually go in too quick, haha, and go for some difficult ones, but try and keep it easy to start with. And then play around at the top end of my ability, falling off a lot, that sort of thing.
\\ LS: Would you class your normal session as training, as fun, or a bit of both ... where on that scale?
\\ P4: Well.. it's definitely all for fun, but i suppose the more you train the more fun it gets, so when I'm there it feels like training, but obviously for fun in the end. I do try to go more regularly to get better and better. So I'm more towards the training than some, I might say.
\\ LS: How has the app impacted your session
\\ P4: Often I don't really think about what I'm doing, I just keep making the same mistakes, and not sure why, so anything that can help break down the climb a bit, even if it's just recording it, like video recording. But then the stats as well, on top of that, are really valuable, just to make me stop and think about what I'm doing more.
\\ LS: So do you use video-recording normally then?
\\ P4: Yeah, I haven't done it a lot, but it is really helpful to get someone to sit there with a camera, but there isn't always someone there.
\\ LS: So that screen with both the video recording and the accelerometer graph, how was that compared to just seeing a video?
\\ P4: Yeah I liked that, I think that yeah the data on its own, I end up mostly just looking at the smoothness in general, but it's hard to work out at which points you were doing those big dynamic moves, but with the video as well, you can see a move that should be very slow and comfortable, if you're getting a big reading from that, then you can work out that that might be contributing to your tiredness.
\\ LS: Ok, cool. So what was your favourite aspect of the app in general?
\\ P4: I do like the smoothness scores, it is a very easy thing to look at, once you know you can do a climb, you do it a few times and try to lower that score as much as possible, it is a very easy way to use the app.
But then the video analysis, if you have someone to record it, that's the most... feature-full bit.
\\ LS: Yeah, so its the mixture of both of those, you have those two different ways of using the app almost.
\\ P4: Yeah exactly, if you're on your own, if I'm just out for a session, and I just want a quick bit of stats, and maybe I know that roughly halfway I was getting some bad results, I;d know roughly where I was going wrong, but if I wanted a full analysis then the video as well would be nice.
\\ LS: Nice. So what was your least favourite aspect of using the app then?
\\ P4: Maybe just like at the start, I wasn't sure what I was able to get from it, I wasn't entirely sure what I should be looking for or doing.
\\ LS: Yeah, so more of an FAQ or some examples of how to use it, on an intro page or something?
\\ P4: Yeah so maybe a breakdown of what you should be looking for in the stats. I worked out that the smoothness was quite important, but maybe some more detail in the instructions there.
\\ LS: I guess that's a downside of me wanting to see how the people testing it would interpret use the numbers, so I didn't want to give exact instructions in case that influenced things...
\\ P4: Yeah, and also I've not been climbing for very long, so don't know the best practices are when it comes to good technique.
\\ LS: So having the app give some advice and rough guidelines for the numbers almost?
\\ P4: Yeah, some people might ignore the advice anyway, but it depends on the level of the climber, personally I want some advice to go with my scores, saying like "this is a bit high", "try to minimise this" and so on
\\ LS: Are you likely to continue using the app in the future
\\ P4: Yeah definitely!
\\ LS: Sound haha! What kind of improvements would you like to see?
\\ P4: I need to figure out the best place to put the phone I think...
\\ LS: Yeah, because you had it in an arm-band for a while didn't you?
\\ P4: Yeah, and I definitely think that has got its uses, like putting it on difficult limbs, to see if you're working harder on each hand for instance, doing the same climb with it on each arm and seeing which is smoother.
\\ LS: That's quite interesting actually!
\\ P4: I didn't hand pockets today, I'll wear pockets next time, I wanna see what results it gives when I've got it attached to my torso.
\\ LS: Yeah, pockets give quite a good approximation for core movement.
\\ P4: Yeah i think it would be less jolty. I definitely want to use the app a few more times to see what kind of data I can get out of it.
\\ LS: Nice. Do you see the app as a training tool, or as a gamification tool?
\\ P4: Probably a training tool yeah...
\\ LS: Haha yeah I guessed that from this conversation we just had
\\ P4: Its good when you've got a lot of people around trying to do the same climb, seeing who can do it the smoothest, I can see how that could be a good game. Today definitely felt more like training though
\\ LS: Yeah, everyone uses the app in a different way, which is kind of why I didn't want to put too much advice on there yet. \\  Do you think the app caused of helped improve your climbing?
\\ P4: Yeah I definitely think that, especially with like rope-climbing or something where it's less powerful and more endurance, I feel I'd get the most use from it there, as I'd be able to use it on climbs where I feel comfortable that I can get most of the way up, but then I'm getting very tired towards the top, it might be helpful to break down how smooth I'm climbing and how much energy I'm able to save using different techniques
\\ LS: So you say those efficiency savings from smooth technique would be more useful in longer rope-climbing routes?
\\ P4: Yeah, its definitely important in bouldering, but you can fall off, have a break and try again, whereas the roped climbs you get more tired on
\\ LS: In bouldering then, which kind of climbs did you mostly use the app with?
\\ P4: Quite often, at the start of a session, when I'm warming up, I'll be looking at my technique, trying to do things slowly and smoothly, like that's when I'd really be looking at the app, then I'll try to remember those smooth movements when I move onto the harder climbs, but sometimes on the really hard ones you are just looking for any way you can get to the top
\\ LS: haha, my sessions are very much the same ha
\\ P4: Yeah so in the warmup stage, its a good chance to practise being smooth and having good technique, the app is good for that.
\\ LS: Are there any features you'd like to see extended or added.
\\ P4: I think it would be great to have external sensors, but I know that would be a massive task ha, connecting to multiple devices isn't easy..
\\ LS: Yeah ha
\\ P4: But the more data you can get, the better
\\ LS: Cool, thanks, any other comments?
\\ P4: Can't think of any
\\ LS: Thank you very much for the interview

\section{Participant Five}
LS: Interview with participant number five, hello!
\\ P5:  Hello
\\ LS: First question, what do your climbing sessions usually involve?
\\ P5:  Usually just a couple of hours, of pretty light climbing if I'm honest, I've just started so haven't done a lot so far.
\\ LS: OK, So would you class those sessions as fun or training, or a mixture?
\\ P5:  Fun
\\ LS: How do you think using the app has impacted your climbing session today?
\\ P5:  I think it helps give me a better awareness of how I was doing on the wall. It's nice to see it displayed on a graph, rather that just trying to remember back to what you did on the wall, gave quite a nice representation of how you'd done that route, and where you could improve.
\\ LS: What would you say was your favourite aspect then?
\\ P5:  The graph visualisation, yeah.
\\ LS: What other features did you use?
\\ P5:  I used the smoothness rating a bit, but because I didn't really have an idea of where the scale went from and to on the rating, the number was a bit meaningless.
\\ LS: Yeah, that makes sense, it can take a few sessions before you get a good feel for what the numbers mean for you. \\ What was your favourite part of using the app?
\\ P5:  It was very user-friendly, I liked the vibrations at the start, to let you know when to climb, that was a nice touch, so you have set it to go, you put it in your pocket, and you know its not recording before you've actually got on the wall.
\\ LS: And your least favourite?
\\ P5:  Probably the fact that you have to keep getting it out of your pocket, rather than just leaving it running for longer, so if I was going to have two of three attempts at the same climb, I could leave it running in my pocket the whole time, then get it out and review all three at the end perhaps.
\\ LS: Yeah, so not having to use the phone as often? Nice idea. Are you likely to use the app in the future?
\\ P5:  Yeah ... I think so, especially if I got more serious about climbing...
\\ LS: You are allowed to say no <laughs>
\\ P5:  <laugh> Yeah, but I think, as a training tool, you know, I think for fun climbers, I don't probably work on a single route enough to get as much real benefit from that app, but if I was training for climbing, and I was doing the same routes a couple of times a week, or a couple of times every session, then having that graph, and the smoothness rating and the data you get, would probably tell you quite a lot about where you were improving
\\ LS: So this is another question actually, do you see the app as more of a training tool then, rather than seeing it as a gamification tool, adding to the fun?
\\ P5:  Yeah, I think I'd use it more as a training tool, rather than adding to any of the fun aspects of climbing
\\ LS: Do you think the app helped or caused any improvements to you climbing?
\\ P5:  definitely when I was using it, it prompted me to think a lot more about how I was climbing, and making sure that I tried to be as smooth as possible on the wall, rather than just lunging for stuff.
\\ LS: So the pressure of the recording...
\\ P5:  Yeah, the pressure of it, at one point I didn't put it on, because I wasn't going for a hold that I knew I could get, because I knew it was in my pocket, recording how much I was moving on the wall, so I stopped...
\\ LS: So is that not always a good thing then, it making you think about smoothness?
\\ P5:  It is a good thing, but having an explanation of the data that the phone is collecting, so like the smoothness rating, having a number, I thought a bit... if it's too high, it's bad, if it's too low it's good, so when I'm climbing I'm just like "gotta get a low number, don't move too much on the wall", that kind of crept into my head. The graph though, that was my favourite bit.
\\ LS: Any features you'd like to see added or extended?
\\ P5:  Probably an expansion of the smoothness rating, so rather than just getting a single number at the end, so like your average score, you could have a bit on the app where you enter a particular route, and then it could produce a scale of expected smoothness rating, for that climber, on that route... 
\\ LS: Ok, sure...
\\ P5:  ..and then you could see how you've progressed over time, or if you've got a dynamic move on that wall, and you want to be really explosive, you can easily compare ratings between different repeats of the climb.
\\ LS: Yeah, there's that scrolling view of climbs currently, so a better...
\\ P5:  Yeah, so almost like being able to group climbs together, and comparing data across different sessions or climbs, rather than just that list of them all individually.
\\ LS: Cool yeah, any further points you'd like to make?
\\ P5:  Nope
\\ LS: Well, thanks very much for you time

\section{Participant Six}
LS:  Interview with participant number 6. hello.
\\ P6: hello
\\ LS:  so 1st question, what do your climbing sessions usually involve?
\\ P6: So I'm trying to improve quite a lot at the moment, improve my upper body strength, really pushing myself, trying to do parts of harder climbs even if I can't do the entire climb.
\\ LS:  So would you class your session as fun or training, or mixture?
\\ P6: A mixture, and depending on how I'm feeling, and depending on how many times I'm climbing in a week, I might do two training sessions and one fun session. Also depending on the people I'm with, if I'm with beginners, I might have more fun doing easier routes with them, but if I'm with people who are better than me then I might push myself a bit more, using their help.
\\ LS:  You've been using the app for a couple of weeks not, how do you think using it has impacted your climbing?
\\ P6: I think it definitely makes you think a bit more when you're on the wall. I tend to be a bit more of a static climber as I'm not so strong as other people that I climb with, so I don't really make the big moves unless I absolutely have to. So yeah, the accelerometer data, there aren't that many big jumps on the graph, it tends to be a bit more steady.
\\ LS:  What does that mean for how you used the app?
\\ P6: I guess I have used the smoothness score thing, but if you compare me to someone with a more dynamic style, then maybe they'd be able to see better where they can make improvements. Because I'm quite a smooth climber anyway, I don't get as much out of that metric.
\\ LS:  Has it been of use in other ways? Can you sometimes see a difference?
\\ P6: Yeah I have still used it because I think it's good to do it over and over again on a climb, and as you get better... for example, if I just jump on a wall, on a route I've not looked at before, and I just set it to go, and then once I know the route, and am climbing it better, yeah I do see an improvement.
\\ LS:  Do you think it's had an impact on your climbing ability, using the app?
\\ P6: Not really no, but it does make me think a bit more when I'm climbing.
\\ LS:  What's your favourite aspect of the app?
\\ P6:L I thought it was really easy to use, just those two big start and stop buttons. It looks quite smart and professional, I like the colours, easy to use I'd say.
\\ LS:  What was your least favourite aspect of the app?
\\ P6: I don't like the fact that it's on a phone. When I climb, I don't have any pockets on my clothing, so I find it quite hard to put it somewhere, I stick it down my shorts and hope it doesn't fall out. I have also used it in a chalkbag, and use it like that, but it bangs around. What I'd like to see in the future is linking it to a wristband or a smartwatch or something...
\\ LS:  So like a lighter weight device separate...
\\ P6: Yeah I think that would be easier for women to use, when we don't have pockets in our clothes.
\\ LS:  Yeah, very true. So do you see the app as a training tool, or more adding more fun to climbing?
\\ P6: For me, I think it's more of a fun thing, but having seen other people use it, I can see how they use it to improve their climbing, and use it in training as well, but for me it's more for fun.
\\ LS:  So could you describe how you'd use it to make the climbs more fun?
\\ P6: I guess it's just interesting, so I could do a climb, and do it really messily, and jumping all over the place, and I see the great big spikes on the graph, and then I could put more thought into it, and be like "oh ok, i could do this", and like, yeah, see the difference.
\\ LS:  So is that on the easier or the harder climbs that you do that?
\\ P6: Yeah, it makes it more fun to climb the easier climbs repeatedly
\\ LS:  Any further points?
\\ P6: No, I think I've covered it all.. Yeah my favourite thing would be to see it in a lighter device, would make it easier to use, although the interface on the app itself is easy, like just some better way of carrying it. I dunno, like if you were to make it more interactive with the wall perhaps, like you can get those walls now that can make the holds light up... Maybe some future app where it can light up the wall, sorry that's really random
\\ LS:  No worries, haha, thanks very much

\end{document}